\documentclass[10pt,reqno]{amsart}

\usepackage{amsmath,amssymb,mathtools}
\usepackage{latexsym}
\usepackage{epsfig}
\usepackage{cite}
\usepackage[english]{babel}
\usepackage{booktabs}
\usepackage{graphicx}
\usepackage{tikz}
\usepackage{pgfplots}
\usepackage{eurosym}
\usepackage{amsthm}
\usepackage{braket}
\usepackage{tikz-feynhand}
\usepackage{caption}
\usepackage{subcaption}

\def\XXint#1#2#3{{\setbox0=\hbox{$#1{#2#3}{\int}$}
     \vcenter{\hbox{$#2#3$}}\kern-.5\wd0}}

\newcommand{\bes}{\begin{equation*}}
\newcommand{\ees}{\end{equation*}}

\numberwithin{equation}{section}

\numberwithin{equation}{section}

\title{Perturbative Approach to Analog Hawking Radiation\\ in dielectric media in subcritical regime}
\author{S. Trevisan$^{3,4}$, F. Belgiorno,$^{1,2}$ S.L. Cacciatori$^{3,4}$ }
\date{}

\address{$^1$ Dipartimento di Matematica, Politecnico di Milano, Piazza Leonardo 32, IT-20133 Milano, Italy}
\address{$^2$ INdAM-GNFM}
\address{$^3$ Department of Science and High Technology, Universit\`a dell'Insubria, Via Valleggio 11, IT-22100 Como, Italy}
\address{$^4$ INFN sezione di Milano, via Celoria 16, IT-20133 Milano, Italy}

\begin{document}

\maketitle
\begin{abstract}
We take into account the subcritical case for dielectric media by exploiting an approximation allowing us to perform perturbative analytical calculations and still not implying low dispersive effects.  
We show that in the background of a  specific soliton-like solution, pair-creation occurs and can display a thermal behaviour governed by an effective temperature.
The robustness of the approach is also corroborated by the analysis of the $\phi\psi$-model related to the standard Hopfield model, for which 
analogous results are obtained.
\end{abstract}

\section{Introduction}

We study the problem of the so-called subcritical case for the Hawking effect in Analogue Gravity, which is characterized by the absence of a real 
horizon but still a process of pair-creation takes place. The phenomenon has been extensively studied in the physical literature, because 
of the Vancouver experiment with water \cite{weinfurtner-prl,weinfurtner-book}, where a thermal spectrum is detected, despite the actual absence of 
a horizon, as shown in subsequent analysis \cite{Parentani_numerical,euve,parentani-sub}.  As to analytical calculations, the first one, again devoted to the case of scattering in water, appeared in \cite{coutant-subcritical}, and it was provided by means of the so-called Bremmer approximation \cite{bremmer49,bremmer}, adapted to an ordinary differential equation of the $n$-th order.  Subsequent analytical calculations, involving also the subcritical case for water and BEC, again in the limit of low dispersive effects, are found in \cite{coutant-kdv,coutant-bdg}, and exploit a KdV equation emerging in the approximation of no co-propagating modes \cite{coutant-kdv}.\\
For the case of the analogue Hawking effect in dielectric media, horizonless situations were numerically taken into account in \cite{fincaru,finazzi-carusotto}, 
but for the subcritical case, a fully analytical calculation is still missing.\\
We have shown in previous papers \cite{master,bec} that 
an equation of the Orr-Sommerfeld type inherited by Nishimoto's works (see e.g. \cite{ni,ni-I}) enables us to treat in a quite unified 
way the case with a horizon, also called the transcritical case, which amounts to the presence of a real turning point (real horizon) in the limit of weak dispersive effects, where the 
weakness of dispersion is indicated by a suitable small parameter $\epsilon$.\\  In this paper, we focus on the subcritical case for dielectric media, i.e. we take into account configurations where a real turning point is missing. We still obtain a fourth-order differential equation governing the phenomenon, 
but we adopt a different attitude and a different expansion parameter with respect to the transcritical case mentioned above. Indeed, we consider 
a linearization of the equation around a specific soliton-like background solution, and we exploit an expansion in terms of a parameter $\eta$ which represents 
the weakness of the soliton amplitude. In the comoving frame of the background solution, we obtain a static situation that mimics that
of the Hawking effect in the transcritical case, but with no horizon. Even if it could seem that weak dispersion is implied by our picture, 
actually this is not the case: Indeed, the dispersion parameter $\epsilon$ is a priori not restricted to be small, and this is the main difference with respect to 
the standard picture described in \cite{master}, where $\epsilon$ is small and a so-called singular perturbation theory is to be allowed. In the 
framework we discuss, one is allowed to adopt a regular perturbative expansion in the parameter $\eta$, which represents a strong advantage.\\
We show that for the particular profile which is taken into account, in some limit thermality is simulated, with an effective temperature which 
is one-third of the one of the corresponding transcritical case.\\
Explicit calculations are firstly carried out for a modified  $\phi\psi$-model 
whose aim is to simplify as possible the dispersion relation associated with the model, and to 
allow more straightforward analytical calculations and a more clear exposition of the basic idea. Then we corroborate the robustness of the 
approach by applying it to the $\phi\psi$ model introduced in \cite{prd2015} and then discussed elsewhere and again taken into account in \cite{master,solitonic}. 
The aforementioned $\phi\psi$ model will be mentioned as the `original' model and represents a simplification of the Hopfield model \cite{hopfield} which is a 
standard way to discuss the electromagnetic field in dielectric media.\\

\section{The modified $\phi\psi$-model and its solitonic solutions}

The model we will consider is defined by a Lagrangian function involving two real scalar fields $\phi$ and $\psi$, with the aim of simulating 
some features of the behavior of the electromagnetic field in dielectric media. As well known, a way to obtain this goal consists in the Hopfield model \cite{hopfield}, 
where the interaction between the electromagnetic field and the atoms/molecules of the dielectric medium is taken into account by replacing the 
aforementioned microscopic objects with a mesoscopic polarization field. One still obtains the correct Sellmeier equation for the dispersion relation. 
In our case, intending to simplify as most as possible both the analytical calculations and the dispersion relation, we replace the polarization field 
with the field $\psi$ and the electromagnetic field with $\phi$, with a set-up aimed at reproducing the Cauchy dispersion relation most straightforwardly. 
We stress that our picture below will be corroborated also in the trickier case of the $\phi\psi$-model discussed in \cite{prd2015}, where the Hopfield model 
is reduced in the most direct way to a model reproducing exactly the Sellmeier dispersion relation using a couple of scalar fields $\phi,\psi$.\\
In our present model, the Lagrangian, expressed in the lab frame, with respect to spacetime variables $t_l,x_l$, is
\begin{align}\label{Eq_Lagrangian}
\mathcal{L} = \frac 1 2 (\partial_{t_l}\phi)^2 + \frac 1 2 \left((\partial_{t_l} \psi)^2 + \mu^2\psi^2\right) + g \phi \partial_{x_l} \psi - \frac \lambda {4!} \psi^4 \,.
\end{align}
We can write it in a covariant form, which will be useful to pass to the frame comoving with the pulse: 
\begin{align}\label{cov_lagrangian}
\mathcal{L} = \frac 1 2 (v^\nu\partial_\nu\phi)^2 + \frac 1 2 \left((v^\nu \partial_\nu \psi)^2 + \mu^2\psi^2\right) + g \phi n^\nu \partial_\nu \psi - \frac \lambda {4!} \psi^4 \,,
\end{align}
where in the lab frame we have $v^\nu_{lab}=(1,0)$ and $n^\nu_{lab}=(0,1)$, while in a boosted frame we get $v^\nu=(\gamma,-V \gamma)$ and $n^\nu=(-V\gamma,\gamma)$ ($c=1$ everywhere). 

The equations of motion that follow from (\ref{cov_lagrangian}) are 
\begin{align}\label{cov-eq}
&(v^\nu\partial_\nu)^2 \phi- g n^\nu \partial_\nu \psi=0,\\
&(v^\nu\partial_\nu)^2 \psi+ g n^\nu \partial_\nu \phi-\mu^2 \phi+\frac{\lambda}{3!} \psi^3=0.
\end{align}
The free-field solutions (for $\lambda=0$) are plane waves $e^{i\omega_{lab} t_l - i k_{lab} x_l}$ which satisfy
\begin{align}\label{CauchyDisp}
n_0^2(\omega_{lab}):=\frac{k_{lab}^2}{\omega_{lab}^2} = \frac {\mu^2}{g^2} + \frac{\omega_{lab}^2}{g^2} =: A + B\omega_{lab}^2 \,,
\end{align}
where $\omega_{lab}= v^\mu k_\mu$ and $k_{lab}= -n^\mu k_\mu$. The quantity $n_0$ is, by definition, the refractive index of the medium. Eq. (\ref{CauchyDisp}), in the limit as $\omega_{lab}\to 0$, gives rise to the Cauchy dispersion relation, which is frequently used to describe dielectric media at low frequencies. Indeed, as $\omega_{lab}\to 0$ we get 
\begin{align}\label{CauchyDisp-v}
n_0(\omega_{lab})\simeq \sqrt{A} + \frac{B}{2 \sqrt{A}}\omega_{lab}^2 \,.
\end{align}
To describe ordinary dielectrics, we require $A>1$, i.e. $\mu>g$.


The Lagrangian (\ref{cov_lagrangian}) admits a conserved current, which is related by the Noether theorem to the invariance (in its complexified version) under phase shifting of the fields $\phi\mapsto e^{i \alpha}\phi$, $\psi\mapsto e^{i \alpha}\psi$:
\begin{align}\label{cov_current}
J^\nu := \frac i 2 \left[ (v^\nu \phi^\ast (v^\alpha\partial_\alpha\phi) +  v^\nu \psi^\ast (v^\alpha\partial_\alpha\psi)  + g n^\nu \psi^\ast \phi- c.c.\right].
\end{align}
The zeroth component gives rise to the conserved charge 
\begin{align}
\int_{\Sigma_t} dx J^0 =: Q,
\end{align}
where $\Sigma_t$ is a spacelike hypersurface. The associated (conserved) scalar product is 
\begin{align}
\left( \begin{pmatrix}
\phi \\ \psi
\end{pmatrix},\begin{pmatrix}
\tilde\phi \\ \tilde\psi
\end{pmatrix}\right)
=  \frac{i}{2} \int_{\Sigma_t} \left[ v^0 \phi^\ast v^\alpha \partial_\alpha \tilde{\phi} - 
v^0 \tilde{\phi} v^\alpha \partial_\alpha \phi^\ast+v^0 \psi^\ast v^\alpha \partial_\alpha \tilde{\psi} - 
 v^0 \tilde{\psi} v^\alpha \partial_\alpha \psi^\ast +  g n^0 (\psi^\ast \tilde{\phi} - \tilde{\psi} \phi^\ast)\right].
\end{align}
It is straightforward now to provide a canonical quantization of the free-lagrangian, compute the propagators and perform standard perturbative QFT computations with the vertex $i\lambda$: we will not deal with this in this paper. The free particles of the theory are \textit{polaritons} which satisfy the Cauchy dispersion relation and the interaction term is a modelization of the nonlinear response that can happen in dielectrics, better known as the Kerr effect. The norm of the free particle states $\Phi_k = \begin{pmatrix}
\phi_k \\\psi_k
\end{pmatrix} \sim e^{-ik_\mu x^\mu}$ has a simple expression in the laboratory frame:
\begin{align}\label{eq_norm}
||\Phi_k||^2 = Q(\phi_k, \psi_k) \propto 2 \omega_{lab}\left(1+ \frac{\omega_{lab}^2}{\mu^2 + \omega_{lab}^2}\right)\,.
\end{align}
From Eq. (\ref{eq_norm}) we can clearly distinguish the positive-norm modes ($\omega_{lab}>0$) from the negative-norm modes ($\omega_{lab}<0$). In Appendix \ref{App_Current} we give a more detailed analysis of the current two-vector $J^\mu$ and we prove that the norm is proportional to $\text{sign}(\omega_{lab})$ in \emph{any} inertial frame. 

The equations of motion (\ref{cov-eq}) also admit a solitonic solution that propagates rigidly at a fixed velocity $V$ with respect to the laboratory:
\begin{align}\label{Eq_Soliton}
\psi_s(x_l-V t_l) &= \frac \alpha {\cosh(\beta (x_l- V t_l))} \,,\\
\alpha^2 &= \frac {12 V^2 \beta^2} {\lambda }\,,\\
\beta^2 &= \frac 1 {V^4} (\mu^2 V^2 - g^2) \,.
\end{align}

\section{The Linearized EOM}

In this section we consider the linearization of the EOMs (\ref{cov-eq}) around a soliton-like background:
\begin{align}\label{Eq_Background}
\psi_B (x_l - V t_l) = \sqrt{\frac{2 |\eta|}{|\lambda|}} \frac 1 {\cosh(\beta (x_l- V t_l))} \,.
\end{align} 
We will treat $\eta$ and $\beta$ as independent parameters to allow valid results not only for the soliton (\ref{Eq_Soliton}) but also for other backgrounds with the same shape. 
Indeed, it is known that solitons of this type can emerge also as \emph{approximated} solutions of electrodynamics inside nonlinear dielectrics: specifically, they are solutions of the so-called nonlinear Schr\"{o}dinger equation, which derives from some approximations made on the Maxwell equations inside those media. 
We shall comment further on our choice in the following section.
 
We want to study the scattering of the asymptotic normal modes against the perturbation given by the soliton travelling across the medium: if the scattering involves both positive and negative-norm modes, this can be interpreted as a sign of instability of the system, which decays by emitting particles, as in the pioneering computation by Hawking for black hole evaporation. 

\begin{figure}
	\begin{subfigure}{.49\textwidth}
		\centering
		\includegraphics[scale=.45]{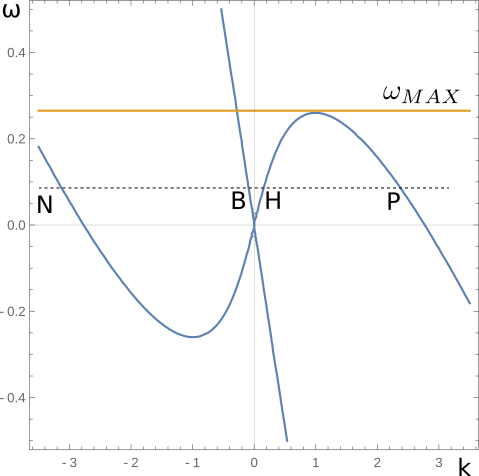}
		\caption{}
		\label{Fig_DR}
	\end{subfigure} \hfill
	\begin{subfigure}{.49\textwidth}
		\centering
		\includegraphics[scale=.45]{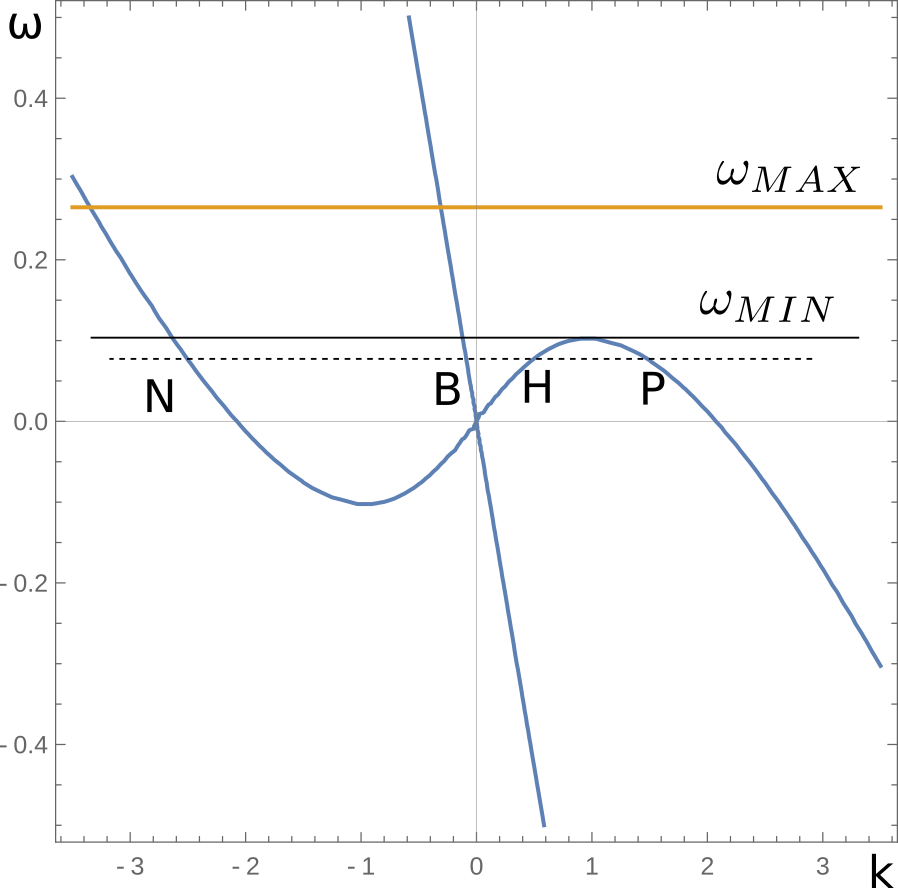}
		\caption{}
		\label{Fig_DRsub}
	\end{subfigure}
\caption{\label{Fig_DR2} (A): The dispersion relation (\ref{CauchyDisp}) represented in the comoving frame with the background, with $g=1$, $\mu=1.2$; for $0<\omega<\omega_{MAX}$ there are four real solutions. (B): The dispersion relation (\ref{Eq_nPerturbed}) represented at the peak of the perturbation, \emph{i.e.} $x_l=V t_l$, for $\eta = -1$. The modes $0<\omega<\omega_{MIN}$ do not experience an event horizon (\emph{subcritical regime}).  }
\end{figure}

The linearized EOMs are
\begin{align} \label{EOM_LinCauchy}
&(v^\nu\partial_\nu)^2 \phi- g n^\nu \partial_\nu \psi=0,\\
&(v^\nu\partial_\nu)^2 \psi+ g n^\nu \partial_\nu \psi-\mu^2 \phi+\frac{\lambda}{2} \psi_B^2 \psi=0. \label{EOM_LinCauchy2}
\end{align}
Notice that, in the linearized equation, the effect of the perturbation is a shift of the parameter $\mu^2$:
\begin{align}\label{Eq_MuShift}
	\mu^2 \mapsto \mu^2 - \frac \lambda {2!}\psi^2_B(x_l-V t_l)\,,
\end{align}
which induces a shift in the refractive index (see (\ref{CauchyDisp})) 
\begin{align}\label{Eq_nPerturbed}
n^2(\omega_{lab},x_l,t_l) &= n_0^2(\omega_{lab}) + \delta n^2(x_l-V t_l)\,, \\
\delta n^2(x_l-V t_l) &= -\frac \eta {g^2}   \frac 1 {\cosh^2(\beta (x_l- V t_l))}\,.
\end{align}
 In order to include both cases $\delta n>0$ and $\delta n < 0$, we will consider $\eta$ as a real quantity, with $$\text{sign}(\eta) :=\text{sign}(\lambda)\,.$$ 
The dispersion relations (\ref{CauchyDisp}) and (\ref{Eq_nPerturbed}) are represented in Figure \ref{Fig_DR2}, as seen in the comoving frame with the background.
 
We decouple the above equations (\ref{EOM_LinCauchy}) and (\ref{EOM_LinCauchy2}), by applying the operator $(v^\nu\partial_\nu)^2 $ to the second equation; then, we eliminate $\phi$ using $(v^\nu\partial_\nu)^2 \phi=g n^\nu \partial_\nu \psi$. In this way, we obtain the following decoupled equation only for the 
field $\psi$:
\begin{align}\label{lin-psi}
(v^\nu\partial_\nu)^4 \psi+ g^2 (n^\nu\partial_\nu)^2 \psi -\mu^2 (v^\nu\partial_\nu)^2\phi+\frac{\lambda}{2} (v^\nu\partial_\nu)^2(\psi_B^2 \psi)=0.
\end{align}
It is convenient to write the equations in the comoving coordinates $t=\gamma (t_l - V x_l)$, $x=\gamma (x_l - V t_l)$. Since the potential term is independent of the comoving time, we seek a solution in the form $\psi = e^{-i\omega t} f(x)$, where the constant $\omega$ is a conserved quantity (cf. e.g. \cite{prd2015}). In this way we end up with a fourth-order equation for $f(x)$ only:
\begin{align}\label{Eq_FourthOrder}
0 =&  V^4 \gamma^4 f^{(4)}(x) +4 i V^3 \omega \gamma^4 f^{(3)}(x) +\gamma^2
f''(x) \left(\eta V^2 \text{sech}^2(\beta x/ \gamma)+g^2-\mu^2 V^2-6 V^2 \omega^2 \gamma^2\right)\cr 
&+2 i V \gamma f'(x) \left(\eta\, \text{sech}^2(\beta x/ \gamma) (2 i \beta V  \tanh (\beta x/ \gamma)+ \omega \gamma)+ \omega \gamma 
\left(g^2-\mu^2-2 \omega^2 \gamma^2\right)\right) \cr 
&+  f(x) \left(-2 \beta^2 \eta V^2 \text{sech}^4(\beta x/ \gamma)-\eta\, \text{sech}^2(\beta x/\gamma) (\omega\gamma +2 i \beta V \tanh (\beta x/\gamma))^2+\omega^2 \gamma^2 \left(-g^2 V^2+ \mu^2+\omega^2 \gamma^2\right)\right)\,.\cr
\end{align}

We can further manipulate the equation by performing the following change of variables, usual for the P\"oschl-Teller potential, 
\begin{align}
z&=-e^{2\tilde\beta x}\,, \quad \tilde \beta := \frac \beta \gamma \,,
\end{align} 
which implies
\begin{align}
\partial_x &= 2\tilde \beta \theta_z := 2\tilde \beta z \frac d {dz}\,.
\end{align}
By defining the rescaled parameters $G=\frac g {2\tilde\beta}$, $\Omega=\frac \omega {2\tilde\beta}$, $M=\frac {\mu} {2\tilde\beta}$, we end up with the following equation
\begin{align}\label{Eq_FuchsianEQ}
0 =& \Big(V^4 \gamma^4 \Big)z^4 f^{(4)} + \Big(6 V^4 \gamma^4  - 4 i V^3 \Omega \gamma^4 \Big) z^3 f^{(3)}\cr 
&+ \left(G^2 \gamma^2 - M^2 V^2 \gamma^2  + 7 V^4 \gamma^4  - 12 i V^3 \Omega \gamma^4  - 
6 V^2 \Omega^2 \gamma^4 -\frac{\eta}{\tilde\beta^2} \frac{ V^2 \gamma^2 z}{(1 - z)^2}\right)z^2 f^{(2)} \cr 
&+\Bigg(G^2 \gamma^2  - M^2 V^2 \gamma^2  - 2 i G^2 V \Omega \gamma^2  + 2 i M^2 V \Omega \gamma^2  + 
V^4 \gamma^4  - 4 i V^3 \Omega \gamma^4  - 6 V^2 \Omega^2 \gamma^4  + 4 i V \Omega^3 \gamma^4 \cr 
&+ \frac{\eta}{\tilde\beta^2}\left(\frac{2 i V \Omega \gamma^2 z}{(1 - z)^2} - \frac{
	V^2 \gamma^2 z (3+z)}{(1 - z)^3}\right) \Bigg)\,z\, f^{(1)}
\cr 
&+ \left(M^2 \Omega^2 \gamma^2 - G^2 V^2 \Omega^2 \gamma^2 + \Omega^4 \gamma^4 +\frac{\eta}{\tilde\beta^2} \left(\frac{\Omega^2 \gamma^2 z}{(1 - z)^2} + \frac{2 i V \Omega \gamma^2 z}{(1 - z)^3} - \frac{ V^2 \gamma^2 z (1+4 z +z^2)}{(1 - z)^4}\right)\right) f  \,.
\end{align} 
Equation (\ref{Eq_FuchsianEQ}) is of Fuchsian type. The complete characterization of the equation is not in the interest of this work and will be the subject of some future publication. In the present work, we propose a perturbative method for the solution of the equation, which will be exposed in the next Section. 

\section{Perturbative Method}\label{sec_VarMethod}

The solitonic solution (\ref{Eq_Soliton}) has no free parameter, except for the velocity $V$: its amplitude, in particular, is fixed by the constants of the Lagrangian. 
Nevertheless, as in the previous section, we linearize the field equations for our nonlinear theory around {\sl weak} soliton-like backgrounds (whichever 
produced experimentally), with the aim to consider the perturbation induced by the soliton-like background as ``small". We stress that we have the freedom to choose the 
spatial dependence of the perturbation in the refractive index in such a way that it can be reproduced using linearization around a given background solution.
This attitude is corroborated  by the experimental fact that 
strong laser pulses 
may induce in dielectrics a non-linear perturbation $\delta n (x)$ (in the comoving frame) of the refractive index which is typically orders of magnitude lower than the leading term 
(e.g. order of $10^{-3}$ compared with one of the leading term).\\
There is an immediate consequence of this approach: as all the non-homogeneity is associated with the refractive index $n(x)$ is due to the correction 
$\delta n(x)$, and the latter one is treated perturbatively in $\eta$ which is associated directly with the amplitude of the 
soliton-like background, we are considering a situation where at the lowest order homogeneity occurs, and then no horizon can appear, i.e. we are 
automatically in a subcritical regime. This is the main difference with respect to the perturbative approach where the perturbation parameter $\epsilon$ is 
associated with a weak dispersion, and where instead a horizon may appear in the leading order equation (in form of a real turning point).

The perturbative method for Fuchsian equations that we apply was proposed in \cite{Chushev}. We start by eliminating the third order derivative by the change of variable $$f(z)=z^{-\frac 3 2 + i \frac \Omega V}u(z)\,,$$ that gives
\begin{align}\label{Eq_full}
u^{(4)} + (v_1(z) + \eta w_1(z))u^{(2)} +(v_2(z) + \eta w_2(z))u^{(1)} + (v_3(z) + \eta w_3(z))u = 0 \,,
\end{align}
where
\begin{align*}
v_1(z) &=\frac{2 G^2 - 2 M^2 V^2 + 5 V^4 \gamma^2}{2 V^4 \gamma^2 z^2} \,,\\
v_2(z) &=\frac{-2 G^2 V + 2 M^2 V^3 + 2 i G^2 \Omega - 2 i G^2 V^2 \Omega - 5 V^5 \gamma^2}{V^5 \gamma^2 z^3} \,,\\
v_3(z) &= \frac{36 G^2 V^2 - 36 M^2 V^4 - 48 i G^2 V \Omega + 48 i G^2 V^3 \Omega - 
	16 G^2 \Omega^2 +32 G^2 V^2 \Omega^2 -16 G^2 V^4 \Omega^2 + 
	81 V^6 \gamma^2}{16 V^6 \gamma^2 z^4} \,,\\
w_1(z) &= -\frac 1 {\tilde\beta^2 V^2 \gamma^2 ( z-1)^2 z} \,,\\
w_2(z) &= \frac 4 {\tilde\beta^2 V^2 \gamma^2 ( z-1)^3 z} \,,\\
w_3(z) &= \frac{-1+2 z-25 z^2} {4 \tilde\beta^2 V^2 \gamma^2 ( z-1)^4 z^3}\,.
\end{align*}
Now we consider $\eta$ as a small parameter, and formally expand the solution as
\begin{align}\label{Eq_FormalSolution}
u(z) = u_{0}(z) + \eta u_{1}(z) + \eta^2 u_{2}(z) + ...
\end{align}
which allows us to obtain a regular perturbative expansion (to be compared with the singular perturbation expansion one obtains by expanding with respect to 
a low dispersion parameter $\epsilon$, cf. section \ref{sec_PhiPsi}). 
Herein, we compute the first order solution $u_1$ and we will discuss the possibility of taking $\eta\rightarrow 1$. 

Solving the unperturbed equation ($\eta=0$) is very easy and it gives 
\begin{align}
u_{0}(z)=z^{i \alpha}\,, 
\end{align}
where $\alpha$ satisfies a fourth-degree algebraic equation.
By putting $i \alpha= i\frac k {2\beta} +\frac 3 2 + i\frac \Omega V$, we find that $k$ is one of the four solutions of the dispersion relation (\ref{CauchyDisp}) as written in the comoving frame:
\begin{align}
\gamma \left(-g^2 (k + V \omega)^2 + (k V + \omega)^2 (\mu^2 + (k V + \omega)^2 \gamma^2)\right)=0\,.
\end{align} 

By substituting (\ref{Eq_FormalSolution}) into (\ref{Eq_full}) we find the set of equations
\begin{align}\label{SetEquations}
u_{1}^{(4)} + v_1(z)u_{1}^{(2)} +v_2(z)u_{1}^{(1)} + v_3(z)u_{1} &= w_1(z)u_{0}^{(2)} +w_2(z)u_{0}^{(1)} + w_3(z)u_{0}\,,\\
&...\,,\cr
u_{n}^{(4)} + v_1(z)u_{n}^{(2)} +v_2(z)u_{n}^{(1)} + v_3(z)u_{n} &= w_1(z)u_{(n-1)}^{(2)} +w_2(z)u_{(n-1)}^{(1)} + w_3(z)u_{(n-1)}
\end{align}
Thus every $u_n$ satisfies a linear differential equation with a source term that depends on $u_{(n-1)}$; the associated homogeneous equation is the unperturbed equation satisfied by $u_0$. 

Let us consider the equation for $u_{1}$. We can explicitly solve it by applying the method of variation of constants. By defining 
\begin{align}\label{Eq_SourceTerm}
r_1(z):= w_1(z)u_{0}^{(2)} +w_2(z)u_{0}^{(1)} + w_3(z)u_{0} \,,
\end{align}
we compute the quantities
\begin{align*}
W(z) &= \det \begin{pmatrix}
z^{i \alpha_1} & z^{i \alpha_2} & z^{i \alpha_3} & z^{i \alpha_4}\\
(z^{i \alpha_1})' & (z^{i \alpha_2})' & (z^{i \alpha_3})' &  (z^{i \alpha_4})'\\
(z^{i \alpha_1})'' & (z^{i \alpha_2})'' & (z^{i \alpha_3})''  & (z^{i \alpha_4})''\\
(z^{i \alpha_1})''' & (z^{i \alpha_2})''' & (z^{i \alpha_3})'''  & (z^{i \alpha_4})'''\\
\end{pmatrix}\,,\\
W_1(z) &= r_1(z) \det\begin{pmatrix}
z^{i \alpha_2} & z^{i \alpha_3} & z^{i \alpha_4} \\
(z^{i \alpha_2})' & (z^{i \alpha_3})'  &  (z^{i \alpha_4})'\\
(z^{i \alpha_2})'' & (z^{i \alpha_3})''  & (z^{i \alpha_4})''\\
\end{pmatrix}\,,\\
W_2(z) &= -r_1(z) \det\begin{pmatrix}
z^{i \alpha_1} & z^{i \alpha_3} & z^{i \alpha_4} \\
(z^{i \alpha_1})' & (z^{i \alpha_3})'  &  (z^{i \alpha_4})'\\
(z^{i \alpha_1})'' & (z^{i \alpha_3})''  & (z^{i \alpha_4})''\\
\end{pmatrix}\,,\\
W_3(z) &= r_1(z) \det\begin{pmatrix}
z^{i \alpha_1} & z^{i \alpha_2} & z^{i \alpha_4} \\
(z^{i \alpha_1})' & (z^{i \alpha_2})'  &  (z^{i \alpha_4})'\\
(z^{i \alpha_1})'' & (z^{i \alpha_2})''  & (z^{i \alpha_4})''\\
\end{pmatrix}\,,\\
W_4(z) &= -r_1(z) \det\begin{pmatrix}
z^{i \alpha_1} & z^{i \alpha_2} & z^{i \alpha_3} \\
(z^{i \alpha_1})' & (z^{i \alpha_2})'  &  (z^{i \alpha_3})'\\
(z^{i \alpha_1})'' & (z^{i \alpha_2})''  & (z^{i \alpha_3})''\\
\end{pmatrix}\,.\\
\end{align*}
A particular solution to the first equation of (\ref{SetEquations}) is given by
\begin{align}\label{Eq_u1_formal}
u_{1}(z) = z^{i\alpha_1}\int dz'\frac{W_1(z')}{W(z')} + z^{i\alpha_2}\int dz'\frac{W_2(z')}{W(z')} + z^{i\alpha_3}\int dz'\frac{W_3(z')}{W(z')} + z^{i\alpha_4}\int dz'\frac{W_4(z')}{W(z')} \,.
\end{align}
In principle, we can recursively iterate the procedure to obtain particular solutions to $u_n$. Notice that the general solution for each order $n$ is obtained by adding a solution to the homogeneous equation, which is the same for every $n$, so we can say that the general solution $u(z)$ is obtained by adding to the iterative solution a combination of the solutions to the unperturbed equation: the coefficients of the combination will be set by the boundary condition of the scattering.

\section{Boundary Conditions for the Scattering}\label{sec_BoundaryConditions}
 
For the subcritical case, it is standard to consider a white hole configuration, in which an initial state, representing a Hawking mode,  approaches the white hole-like perturbation at early times, and four modes emerge at late times: three backward modes $P,N,B$, where $N$ is the only negative norm mode appearing in the scattering, and a transmitted-mode $T$ representing the fraction of the Hawking mode $H$ which is transmitted beyond the perturbation This is nothing but 
what happens in presence of a real white hole horizon in the transcritical case, apart for the transmitted mode. For the black hole case, 
an analytical study is trickier, as the black hole-like configuration, in principle, is not related to the white hole-like one by time reversal: in presence of a black hole horizon, one has three entering initial modes $P,N,B$ which are converted in a scattered emerging mode $H$. When the scattering is subcritical, in principle, given three initial modes $P,N,B$, one should consider the possibility to get three transmitted particles, and then it is evident that one does not obtain the time reversal configuration of a  white hole subcritical scattering. 

We consider a white hole-like configuration, as usual. Thus, in the initial state, we will have only a right-moving $H$-mode, which is scattered and mode converted in a way such that we obtain the four modes in the final state.
The asymptotic form of the solution we seek is thus
\begin{align}\label{Eq_AsymptoticBehaviour}
f(x) \sim \begin{cases}
e^{i k_H x} + C_2 e^{i k_P x} +C_3 e^{i k_N x}+C_4 e^{i k_B x}  \,,\quad & x\rightarrow -\infty \\
 C_1 e^{i k_H x}   \,,\quad & x\rightarrow +\infty
\end{cases}\,,
\end{align}
when we recall that the $H$-mode has a positive group velocity (right-moving), while all other modes have negative group velocity (left-moving). Since the unperturbed solution would be just $f(x)=e^{i k_H x}$ defined everywhere, we expect the coefficient $C_1$  of the transmitted part of the Hawking mode to be $O(1)$, while all other coefficients must be $O(\varepsilon)$. We can now proceed to the solution of the equation (\ref{SetEquations}) for $u_1(z)$ considering for the source term the unperturbed solution $u_0(z)=z^{i \alpha_H}$, where $i \alpha_H= i\frac {k_H} {2\beta} +\frac 3 2 + i\frac \Omega V$. The detailed computation of the solution is shown in Appendix \ref{app_Solution}: in what follows we will just expose the final expressions.

The asymptotic coefficients of the solution (see Eq. (\ref{Eq_AsymptoticBehaviour})) at first perturbative order are
\begin{align}\label{Eq_Coefficients}
C_1&= \frac{1}{1+i \eta \tau}\,, \\ 
C_2&= \frac{i\eta}{1+i \eta \tau}\,\frac{ \pi (k_P V + \omega)^2 (-1)^{i (k_H - k_P)}
	}{4\beta^2 V^4 (k_P - k_N) (k_P - k_B) }\mathrm{csch}\left((k_H - k_P) \frac {\pi\gamma} {2\beta}\right)\,, \\
C_3&= \frac{i\eta}{1+i \eta \tau}\, \frac{ \pi (k_N V + \omega)^2 (-1)^{i (k_H - k_N)}
	}{4\beta^2 V^4 (k_N - k_P) (k_N - k_B) }\mathrm{csch}\left((k_H - k_N) \frac{\pi\gamma}{2\beta}\right)\,, \\
C_4&= \frac{i\eta}{1+i \eta \tau}\, \frac{ \pi (k_B V + \omega)^2 (-1)^{i (k_H - k_B)}
	}{ 4\beta^2 V^4 (k_B - k_P) (k_B - k_N) }\mathrm{csch}\left((k_H - k_B) \frac{\pi\gamma}{2\beta}\right)\,,\\
\tau &= \frac{ (k_H V + \omega)^2}{2\beta V^4 \gamma(k_H - k_P) (k_H - k_N) (k_H - k_B) }  \,.
\end{align}

\section{Results}

Current conservation implies that the following equation holds true: 
\begin{equation}
|J_H|=|J_T| + |J_P|-|J_N|+|J_B|, 
\end{equation}
so that, by defining
\begin{align}
|T|&=\frac{|J_T|}{|J_H|},\\
|N|&=\frac{|J_N|}{|J_H|},\\
|B|&=\frac{|J_B|}{|J_H|},
\end{align}
we get 
\begin{align}\label{Eq_CurrentConservation}
1  = |T| +  |P| - |N| + |B| \,,
\end{align}
where  for the model labeled by $K$  we have $|K|:=|C_K|^2 \left|\frac{v_K\partial_\omega DR|_{k_K}}{v_H\partial_\omega DR|_{k_H}}\right|$, $v_K(\omega)$ is the group velocity of the $K$-mode and $\text{DR}(\omega, k)$ is the dispersion relation function defined in Eq. (\ref{Eq_DR}): we provide a derivation of (\ref{Eq_CurrentConservation}) in Appendix \ref{App_ConservationLaw}. The ratio $|N|$ corresponds, in the black hole case, to the rate of spontaneous emission of $H$-mode waves, as it can be argued by using standard Bogoliubov coefficients. In the present case, it represents the rate of pair-production 
in the subcritical process at hand.\\
From the scattering coefficients (\ref{Eq_Coefficients}) and the expression (\ref{Eq_smartfluxes}) for the flux factors, we can write down explicitly the expressions
\begin{align}
|N|&= \frac {\eta^2} {1 + \eta^2 \tau^2}\frac{ \pi^2 (k_N V + \omega)^2 (k_H V + \omega)^2 \mathrm{csch}[((k_H - k_N)\gamma\pi)/(
2 \beta)]^2}{16\beta^4 V^8 (k_N - k_P) (k_N - k_B)(k_H - k_P) (k_H - k_B)  } \label{Eq_N}\\
\frac{|P|}{|N|} &= \frac{(k_N - k_B)(k_P-k_H) (k_P V + \omega)^2 (\mathrm{csch}[\pi\gamma(k_H - k_P)/(2\beta)])^2}{(k_P - 
	k_B)(k_N-k_H) (k_N V + \omega)^2 (\mathrm{csch}[\pi\gamma(k_H - k_N)/(2\beta) ])^2}  \,, \label{Eq_PoN}\\
|T| &= \frac 1 {1 + \eta^2 \tau^2}\,, \label{Eq_H} \\
|B| & = \frac {\eta^2} {1 + \eta^2 \tau^2}\frac{\pi^2  (k_B V + \omega)^2 (k_H V + \omega)^2  (\mathrm{csch}[\pi\gamma(k_H - k_B)/(2\beta) ])^2} {16\beta^4 V^8(k_B - k_N) (k_B - k_P)(k_H-k_N)(k_H-k_P)  } \,. \label{Eq_B} 
\end{align}
As a consequence, we can provide 
\begin{align}
|N| = \frac{1-|T|-|B|}{\frac{|P|}{|N|}-1} \,.
\end{align}\
We stress that these expressions are exact except for the perturbative approximation in $\eta$: no other approximation has been made throughout the computation, so they are valid for all frequencies, as long as we know $k_j(\omega)$.

\begin{figure}
	\begin{subfigure}{.49\textwidth}
		\centering
		\includegraphics[scale=.7]{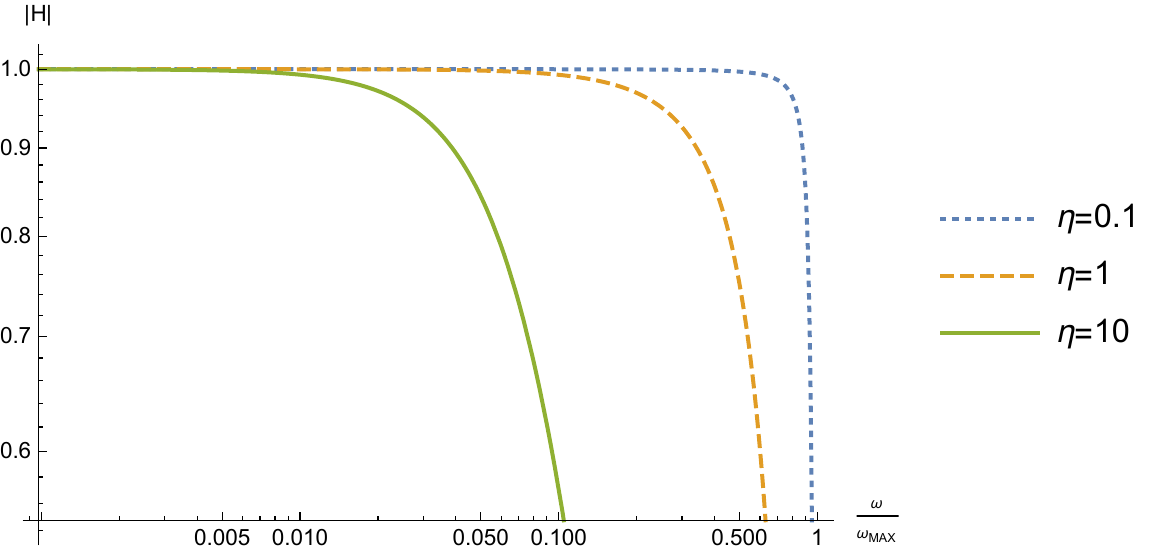}
		\caption{}
		\label{Fig_HofL}
	\end{subfigure} \hfill
	\begin{subfigure}{.49\textwidth}
	\centering
	\includegraphics[scale=.7]{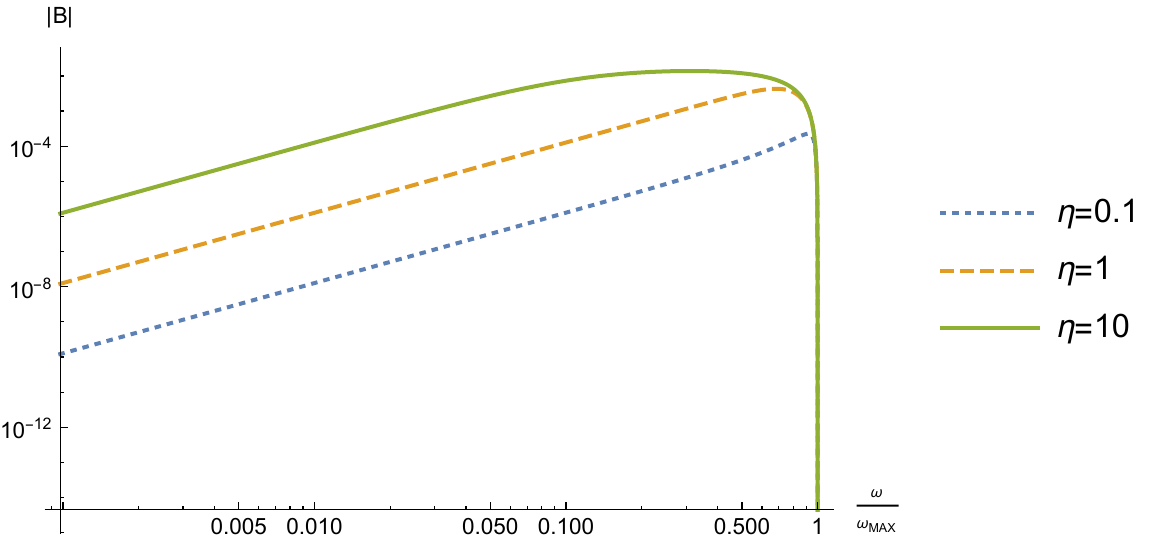}
	\caption{}
	\label{Fig_BofL}
\end{subfigure}
\caption{The coefficients $T$ and $B$ from (\ref{Eq_H}) and (\ref{Eq_B}), for different values of the amplitude of the perturbation $\eta$. }
\label{Fig_HeB}
\end{figure}

\begin{figure}
	\begin{subfigure}{.49\textwidth}
		\centering
		\includegraphics[scale=.7]{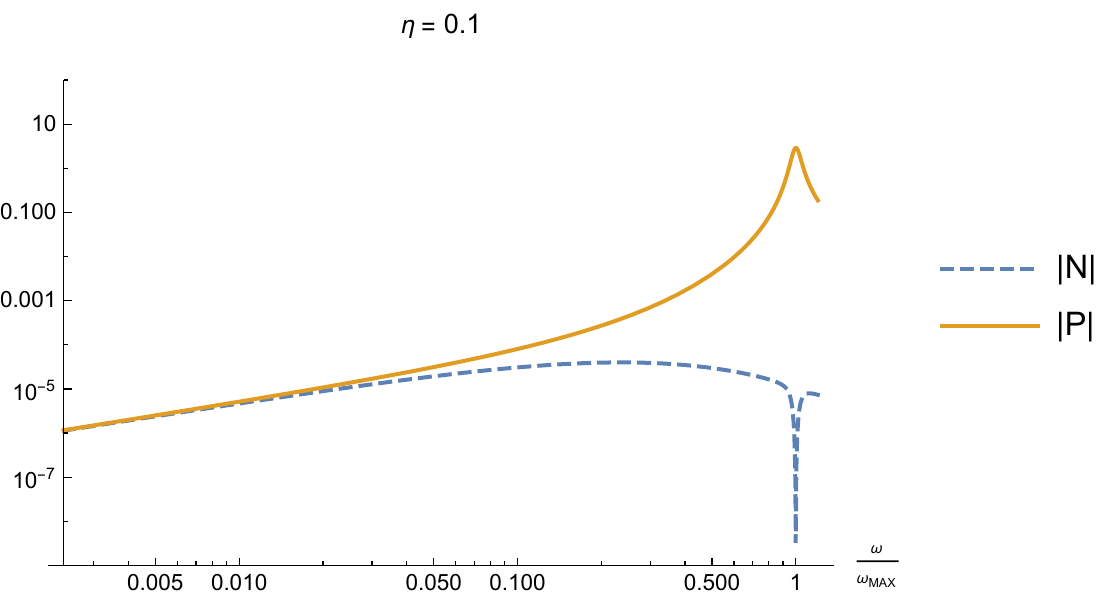}
		\caption{}
		\label{Fig_PNe01}
	\end{subfigure} \hfill
	\begin{subfigure}{.49\textwidth}
		\centering
		\includegraphics[scale=.7]{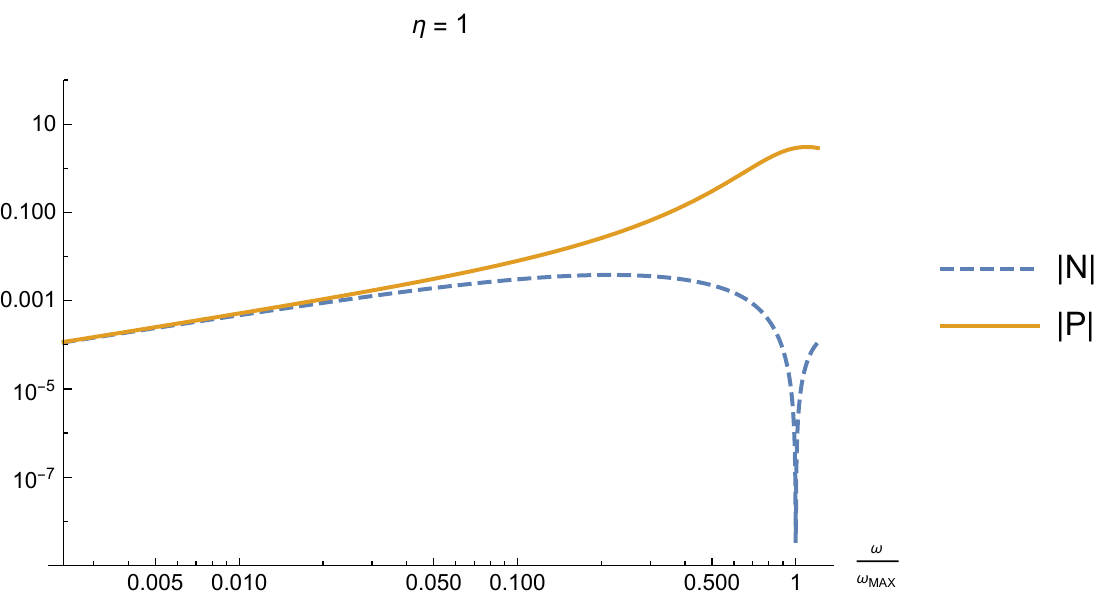}
		\caption{}
		\label{Fig_PNe1}
	\end{subfigure}\\
\vspace{4mm}
\begin{subfigure}{.49\textwidth}
	\centering
	\includegraphics[scale=.7]{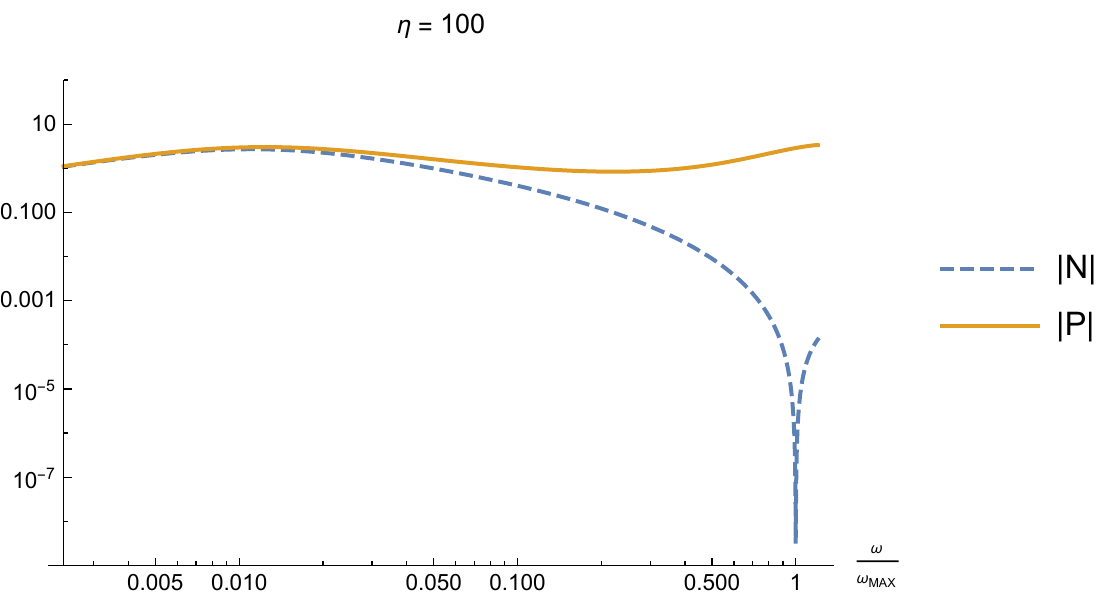}
	\caption{}
	\label{Fig_PNe100}
\end{subfigure} \hfill
\begin{subfigure}{.49\textwidth}
	\centering
	\includegraphics[scale=.7]{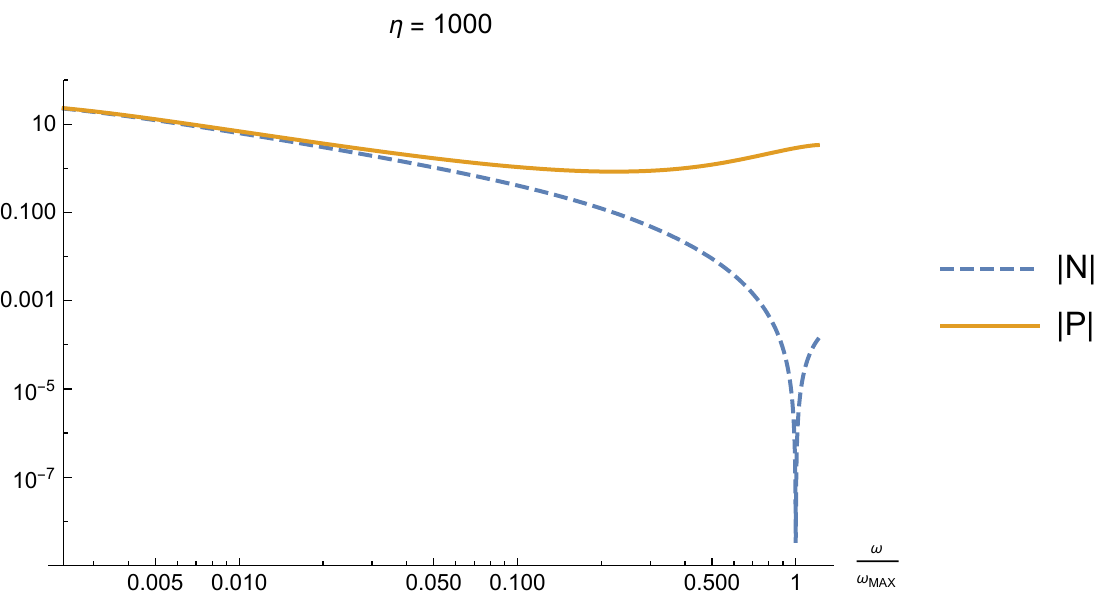}
	\caption{}
	\label{Fig_PNe1000}
\end{subfigure}
	\caption{The coefficients $P$ and $N$ from (\ref{Eq_N}) and (\ref{Eq_PoN}), for different values of the amplitude of the perturbation $\eta$. In Figures (A)-(C) the behaviour at small frequencies is $P, N \sim \omega$, which agrees with the findings of \cite{coutant-subcritical}; in Figure (C) we show the onset of critical behaviour, as $P,N\sim \omega^{-1}$. }
	\label{Fig_PN}
\end{figure}

\begin{figure}
	\begin{subfigure}{.49\textwidth}
		\centering
		\includegraphics[scale=.7]{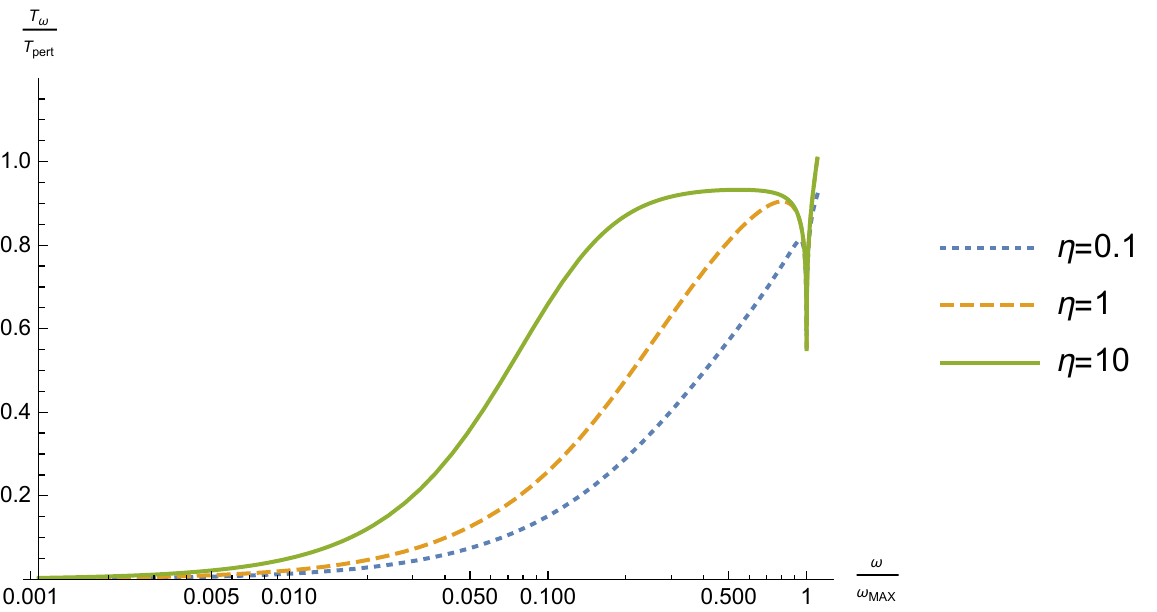}
		\caption{}
		\label{Fig_TofL}
	\end{subfigure} \hfill
	\begin{subfigure}{.49\textwidth}
		\centering
		\includegraphics[scale=.7]{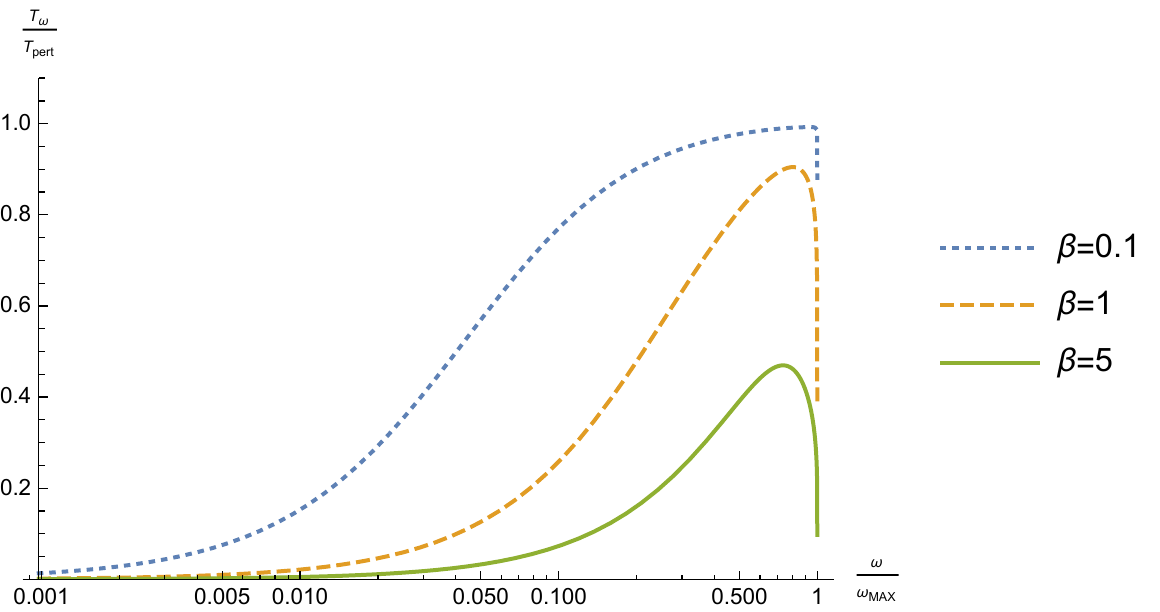}
		\caption{}
		\label{Fig_Tofb}
	\end{subfigure}
	\caption{The temperature $T_\omega$ as defined in (\ref{Eq_Tw}). (A): Plot at fixed $\beta=1$, for different $\eta$; we see the formation of a plateau at $T_\omega \approx T_{pert}$ for high $\eta$. (B): Plot at fixed $\eta = 1$, for different $\beta$; the estimation of the value of the plateau $T_{pert}$ is less accurate for increasing $\beta$. }
	\label{Fig_Tw}
\end{figure}

\begin{figure}
	\centering
	\includegraphics[scale=.7]{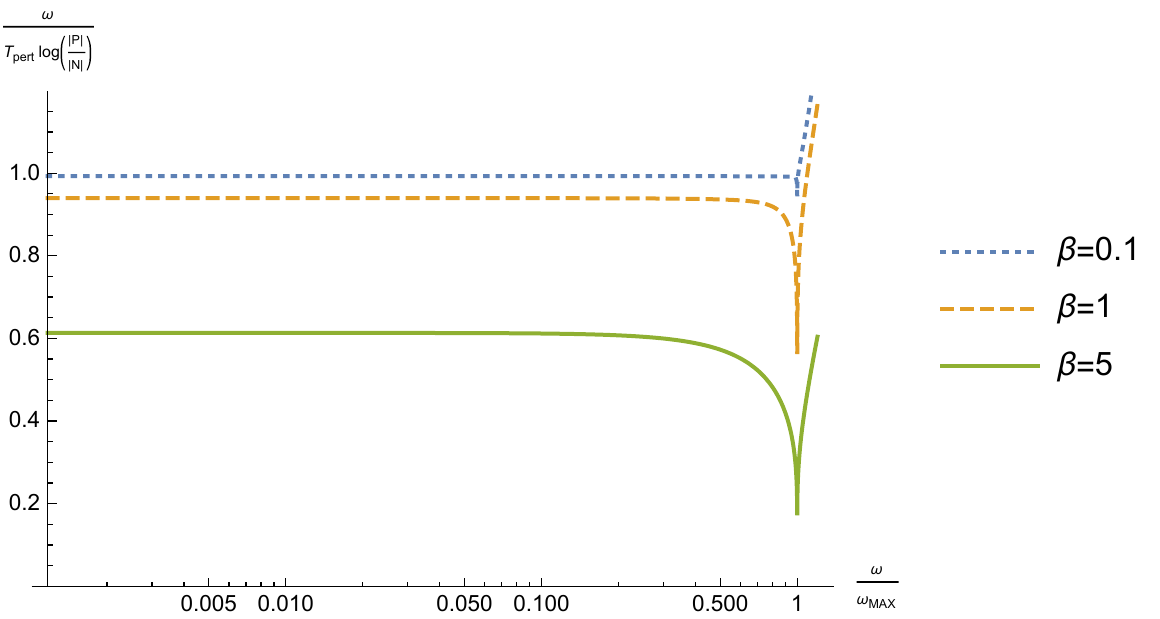}
	\caption{The ratio $ \omega / {\log(\frac{|P|}{|N|})}$ normalized by the estimated temperature $T_{pert}$ (see Eq. (\ref{Eq_TH})), as a function of $\beta$. The ratio is almost constant, which means that $\frac{|P|}{|N|}$ is indeed exponential with the frequency; our estimation of the "temperature" $T_{pert}$ proves to be good at low $\beta$, while less accurate for increasing $\beta$. }
	\label{Fig_LogPoN}
\end{figure}

For an explicit evaluation we have expanded the modes $k_j(\omega)$ from Eq. (\ref{CauchyDisp}) near $\omega=0$, obtaining
\begin{align}
k_H &= \frac{\mu - g V}{g - \mu V}\,\omega + O(\omega^3), \\
k_B &= -\frac{\mu + g V}{g + \mu V}\,\omega + O(\omega^3), \\
k_P &=  \frac{\sqrt{g^2 - \mu^2 V^2}}{
	\gamma V^2 } - \left(\frac 1 V + \frac {g^2} {\gamma^2 V (g^2 - \mu^2 V^2)}\right) \,\omega  - \frac{g^2 (2 g^2 + \mu^2 V^2) }{2\gamma  (g^2 - \mu^2 V^2)^{5/2} }\,\omega^2 + O(\omega^3) , \\
k_N &=  -\frac{\sqrt{g^2 - \mu^2 V^2}}{
	\gamma V^2 } - \left(\frac 1 V + \frac {g^2} {\gamma^2 V (g^2 - \mu^2 V^2)}\right) \,\omega  + \frac{g^2 (2 g^2 + \mu^2 V^2) }{2\gamma  (g^2 - \mu^2 V^2)^{5/2} }\,\omega^2 + O(\omega^3)\,,
\end{align}
and we put $g=1$, $\mu=1.2$, $V=0.5$. As to $k_T$, due to the fact that the perturbation vanishes very rapidly in the comoving frame both for $x\to \infty$ and for  
$x\to -\infty$, one easily find $k_T=k_H$. 
These values are taken just to obtain qualitatively the same dispersion relation as in an experimental situation of laser pulses in silica, but they are not meant to be quantitatively accurate. In the following analysis of the results we will treat both $\eta$ and $\beta$ as independent parameters: although they are uniquely determined by the soliton (\ref{Eq_Soliton}), we still think that it is interesting to study the dependence of the radiation on these parameters. We might equivalently say that we are studying a generic perturbation of the form (\ref{Eq_Background}) with generic parameters, even though this perturbation might not be a solution of the nonlinear equations of motion (\ref{cov-eq}). 
The scattering coefficients $|T|$, $|B|$, $|P|$ and $|N|$ are plotted in Figure \ref{Fig_HeB} and Figure \ref{Fig_PN}, for different values of $\eta$. Notice that $|B|$ results to be smaller than the others, as was expected. The transmission coefficient $|T|$ is $\sim 1$ for low frequencies but it decreases more rapidly as $\eta$ increases: this is a sign that for high perturbation we approximate the formation of an event horizon: this interpretation is supported by the fact that the frequencies near $\omega_{MAX}$ are less transmitted. 

The coefficients $|P|$ and $|N|$ are nearly equal at low frequencies and are $\sim \omega$. This behaviour is in agreement with what was found by Coutant and Weinfurtner \cite{coutant-subcritical} for a subcritical flow in shallow water. 

The ratio $\frac {|P|}{|N|}$ is very close to an exponential function, $\frac {|P|}{|N|}\sim e^{\frac{\omega}{T}}$. This in shown if Figure \ref{Fig_LogPoN}, where we see that $\omega / \log(\frac{|P|}{|N|})$ is almost constant. The ``temperature" $T_{pert}$ (the subscript is for ``perturbative'') can be estimated analytically from the arguments of the $\mathrm{csch}$ in (\ref{Eq_PoN}): we find
\begin{align}\label{Eq_TH}
\beta_{pert} = \frac 1 {T_{pert}} = \frac{2\pi \gamma}{2\beta} \lim\limits_{\omega\rightarrow 0} \frac{2 k_H - k_P - k_N}{\omega} = \frac {2\pi g (2 g + \mu V)  }{ \beta  \gamma V (g^2 - \mu^2 V^2)}  \,.
\end{align}
We note that $T_{pert}\propto\beta$, which is proportional to the derivative of the background function (\ref{Eq_Soliton}): this is what was expected based on previous literature about analogue Hawking radiation in the critical case. As it is clear from Figure \ref{Fig_LogPoN}, while the exponential approximation is still valid, our estimation of the temperature is less good as $\beta$ increases.

Finally, we define a temperature function $T_\omega$ by the relation
\begin{align}\label{Eq_Tw}
|N|=:\frac 1 {e^{\frac \omega {T_\omega}}-1}\,.
\end{align}
The function $T_\omega$ is plotted in Figure \ref{Fig_Tw} for different values of $\eta$ and $\beta$. We note $T_\omega$ is not constant, showing a lack of thermality in the emission spectrum. However, for higher values of $\eta$, a plateau is created for frequencies close to $\omega_{MAX}$, with $T_\omega\approx T_H$. This fact is again interpreted as a sign of the presence of a group horizon for higher frequencies: indeed, a similar behaviour was found by a numerical study of the transcritical regime in shallow water \cite{Parentani_numerical}. 

The near criticality of the system for high $\eta$ is underlined also by the behaviour of $|P|$ and $|N|$, which start growing as $\sim \frac 1 \omega$ near $\omega=0$ (see Figure \ref{Fig_PNe1000}). This is precisely the behaviour that one expects in the critical case. We specify that the results for high values of $\eta$ should be taken carefully since $\eta$ is precisely the expansion parameter of our perturbative solution. However we find it interesting that already at the leading order, the solution shows a near critical behaviour for high $\eta$; we expect that the same qualitative behaviour is present also in the higher order solutions, possibly showing up already at lower values of $\eta$.

\section{The original $\phi\psi$-model}\label{sec_PhiPsi}

In the original $\phi\psi$-model, including a $\psi^4$ term, one has 
\begin{align}
{\mathcal L}_{\varphi\psi} = \frac{1}{2} (\partial_\mu \phi)(\partial^\mu \phi)+\frac{1}{2\chi \omega_0^2} \left[ (v^\alpha \partial_\alpha \psi)^2 - \omega_0^2 \psi^2 \right] + \frac{1}{c} (v^\alpha \partial_\alpha \psi) \phi-\frac{\lambda}{4!} \psi^4,
\label{Lagrangian-diele}
\end{align}
where $\phi,\psi$ play the role of the electromagnetic field and polarization field respectively, $\chi$ plays the role of the dielectric susceptibility, $v^\mu$ is the usual four-velocity vector of the dielectric, $\omega_0$ is the proper frequency of the medium \cite{solitonic}. We get 
the system
\begin{align}
\label{phi-eq}
\Box \phi - \frac{1}{c} (v^\mu \partial_\mu \psi) & =0, \\
\label{psi-eq}
\left(\frac{1}{\chi \omega_0^2} (v^\mu \partial_\mu)^2 +\frac{1}{\chi}\right) \psi+  
\frac{1}{c} (v^\mu \partial_\mu \phi)+ \frac{\lambda}{3!} \psi^3& =0 . 
\end{align}
As in~\cite{hopfield-kerr}, we allow  the spatial dependence to appear in $\chi$ and in $\omega_0$ 
in such a way that $\chi \omega_0^2=$ const. Indeed, by linearizing the model around soliton-like solutions, directly written in the comoving frame, one obtains 
\begin{eqnarray}
\frac{1}{\chi} &\mapsto& \frac{1}{\chi} + \frac{\lambda}{2} \psi_0^2 (x),\\
\omega_0^2 &\mapsto& \omega_0^2  (1+\chi \frac{\lambda}{2} \psi_0^2 (x)), 
\end{eqnarray}
in such a way that $\chi \omega_0^2 $ remains invariant: $\chi \omega_0^2 \mapsto \chi \omega_0^2 $.\\

In this case, we identify the parameter associated with dispersion as follows: 
$$
\epsilon^2 \coloneqq \frac{1}{\chi \omega_0^2} .
$$
This parameter has been considered as the small parameter associated with the model in the limit 
of low dispersion.\\
In the present case, we adopt a different view where dispersion can also be strong, and the expansion parameter 
is instead associated with the amplitude of the background solution around which the EOMs are linearized. We limit ourselves to notice that 
a linearization of the EOMs around a background soliton-like solution $\psi_0 (x)$ amounts simply in replacing $\frac{\lambda}{3!} \psi^3\mapsto 
\frac{\lambda}{2} \psi_0^2 \psi$ in \eqref{psi-eq}. 

\subsection{A separated equation for $\psi$}

By  applying the operator $\Box$ on the left of equation~\eqref{psi-eq}, as shown in \cite{master}, we obtain the following fourth-order ordinary differential 
equation:
\begin{equation}
\begin{split}
&-\epsilon^2 \partial_x^4 f - 2 i \epsilon^2 \frac{\omega}{v} \partial_x^3 f+ \frac{1}{\chi \gamma^2 v^2}
\left( -\biggl(1- \chi \gamma^2 \frac{v^2}{c^2}\biggr) +\epsilon^2 \chi \omega^2\right) \partial_x^2 f+
2 \biggl(i \frac{\omega}{v} \frac{1}{c^2} (1 - \epsilon^2 \omega^2) - \frac{1}{\gamma^2 v^2} \biggl(\partial_x \frac{1}{\chi}\biggr) \biggr) \partial_x f \\
& + \left( \epsilon^2 \frac{\omega^4}{v^2 c^2}-   \frac{1}{\gamma^2 v^2} \biggl(\partial_x^2 \frac{1}{\chi}\biggr)
- \frac{\omega^2}{\chi\gamma^2 v^2 c^2} - \frac{\omega^2}{c^2 v^2} \right) f =0.
\end{split}
\end{equation}
We also define $f(x) = h(x) \zeta(x)$, with
\begin{equation}
h(x) = A \exp (-i \frac{\omega}{2 v} x),
\end{equation}
where $A$ is a constant. $h(x)$ is chosen such that the third-order term vanishes, and the procedure is analogous to 
the Liouville transformation which eliminates the first-order term in a second-order linear ordinary differential equation. 
This leads to the following quartic equation, which is just of the type `Orr--Sommerfeld' 
\begin{equation}
\label{orso-psi}
\begin{split}
&-\epsilon^2 \partial_x^4 \zeta + \biggl[ -\frac{1}{\chi \gamma^2 v^2} \biggl(1- \chi \gamma^2 \frac{v^2}{c^2}\biggr) 
+\epsilon^2 \frac{1}{\gamma^2 v^2}  \biggl(1-\frac{3}{2} \gamma^2\biggr) \omega^2\biggr] \partial_x^2 \zeta   \\
& +\biggl(  i \frac{\omega}{v} 
\frac{1}{\chi \gamma^2 v^2} 
\biggl( 1+\chi \gamma^2 \frac{v^2}{c^2} \biggr) -2 \frac{1}{\gamma^2 v^2} \biggl(\partial_x \frac{1}{\chi}\biggr) -i \epsilon^2 \frac{\omega^3}{v c^2}\biggr) \partial_x \zeta \\
& +
\biggl[\frac{1}{ \gamma^2 v^2} \biggl( i \frac{\omega}{v} \biggl(\partial_x \frac{1}{\chi}\biggr) -\biggr(\partial_x^2 \frac{1}{\chi}\biggr)\biggr) 
+ \frac{1}{ \gamma^2 v^2} \biggl( \frac{1}{4 \chi} \frac{\omega^2}{v^2} \biggl(1-\chi \gamma^2 \frac{v^2}{c^2} \biggr) 
-\frac{\omega^2}{\chi c^2} \biggl) \\
&
+\epsilon^2 \biggl(\frac{\omega^4}{v^4} \biggl(-\frac{1}{16}+\frac{1 v^2}{4 c^2}\biggr)\biggr) \biggr] \zeta =0.
\end{split}
\end{equation} 
The effect of the linearization around a background soliton-like solution $\psi_0 (x)$ consists simply in the 
replacement
\begin{eqnarray}
\frac{1}{\chi} \mapsto\frac{1}{\chi_0} + \frac{\lambda}{2} \psi_0^2 (x), 
\end{eqnarray}
where $\chi_0$ is constant. We choose 
\begin{align}
\psi_0 (x):= 2 \sqrt{\frac{|\eta|}{|\lambda|}} \frac{1}{\cosh (\tilde\beta x)}.
\end{align} 
As in the model discussed in the previous sections, we turn to the change of variable 
\begin{align}
z:=-\exp(2 \tilde\beta x).
\end{align} 
As a consequence, with some abuse of language, we have $\zeta = f(z)$. 
Then we obtain 
\begin{eqnarray}
\frac{1}{\chi} \mapsto\frac{1}{\chi_0} - 2 \eta \frac{z}{(1-z)^2}. 
\end{eqnarray}
Furthermore, to cancel a new third-order term that appears again after the above-mentioned independent variable change, we set 
$\zeta(z) =z^{-3/2} h(z)$. Then we obtain the following equation: 
\begin{align}\label{eq_FinalEquation_phipsi}
h^{(4)} + (u_2+\eta w_2) h^{(2)}+ (u_1+\eta w_1) h^{(1)}+(u_0+\eta w_0) h=0, 
\end{align} 
where 
\begin{align*}
u_2 &=\frac{\frac{1}{\gamma^2 v^2 \chi_0}(1-\chi_0 \gamma^2 \frac{v^2}{c^2})+\epsilon^2 (10 \tilde\beta^2  -(1-\frac{3}{2} \gamma^2)\omega^2) }{4 \tilde\beta^2 \epsilon^2 z^2} \,,\\
u_1 &=\frac{-4\tilde\beta v c^2 (1-\chi_0 \gamma^2 \frac{v^2}{c^2})-i c^2 \omega (1+\chi_0 \gamma^2 \frac{v^2}{c^2})+\epsilon^2 \chi_0 v (-40 \tilde\beta^3 v^2 c^2 \gamma^2+2 \tilde\beta c^2 (2-3 \gamma^2) \omega^2 +i \gamma^2 v \omega^3 )}{8 \tilde\beta^3 c^2 \chi_0 \epsilon^2 \gamma^2 v^3 z^3} \,,\\
u_0 &= \frac{144 \tilde\beta^2 v^2 c^2 (1-\chi_0 \gamma^2 \frac{v^2}{c^2})+48 i \tilde\beta v c^2 (1+\chi_0 \gamma^2 \frac{v^2}{c^2})+4 \omega^2 ((4+\chi_0 \gamma^2)v^2-c^2)+ \epsilon^2 \delta}{256 \tilde\beta^4 c^2 \chi_0 \epsilon^2 \gamma^2 v^4 z^4} \,,\\
w_2 &= -\frac{1}{2\tilde\beta^2 \epsilon^2 \gamma^2 v^2 z (1-z)^2} \,,\\
w_1 &= \frac{i \omega (z-1)+8 \tilde\beta v z}{4\tilde\beta^3 \epsilon^2 \gamma^2 v^3 z^2 (-1+z)^3} \,,\\
w_0 &= \frac{(c^2-4 v^2) \omega^2 (z-1)^2+4 i \tilde\beta c^2 v \omega (1-z)(5 z-1)-4 \tilde\beta^2 c^2 v^2 (1-2 z+25 z^2)} {32\tilde\beta^4 c^2 \epsilon^2 \gamma^2 v^4 ( z-1)^4 z^3}\,,
\end{align*}
with 
$$
\delta:= \chi_0 (1296 \tilde\beta^4 v^4 c^2 \gamma^2 +72 \tilde\beta^2 c^2 v^2 \omega^2 (-2+3\gamma^2) -48 i \tilde\beta \gamma^2 v^3 \omega^3 +\gamma^2 (c^2-4 v^2) \omega^4).
$$

\subsection{Dispersion relation and its roots}

As in the previous case, we can guess that solutions of the zeroth order equation are of the form $z^{i \alpha}$ and that $\alpha$ satisfies the 
dispersion relation associated with the model. In the comoving frame, the eikonal equation for the model provides us with the following equation: 
\begin{equation}
\left(k^2-\frac{\omega^2}{c^2}\right) \frac{1}{\chi \omega_0^2} (\omega_0^2-\gamma^2 (\omega+ v k)^2) -\frac{1}{c^2} \gamma^2 (\omega+ v k)^2=0. 
\label{disp-fipsior}
\end{equation}
It is a quartic equation whose roots $k_i (\omega)$, $i=1,2,3,4$ cannot be managed in simple formulas unless some kind of approximation is provided. 
Our ansatz is the following. We choose $\delta \to 0$ as an expansion parameter and we put 
\begin{align}
k &=: \frac{1}{\delta} u-\frac{\omega}{v},\\
\omega_0 &=: \frac{1}{\delta} \bar{\omega}_0. 
\end{align} 
We mean to indicate an expansion where $k$, unless it is zero, dominates over $\omega/v$, i.e. $\omega$ is small relative to a nonzero $k$, and that 
also $\omega_0$ is big compared to $\omega$ (this kind of approach leads also to the Cauchy approximation, cf. e.g. \cite{hawbook}, ch. 9). As a consequence, we find 
the following re-writing of \eqref{disp-fipsior}:
\begin{equation}
\left(u^2-2 \delta \frac{\omega}{v} u+\delta^2 \frac{\omega^2}{\gamma^2 v^2}\right)\left(1-\frac{\gamma^2 v^2}{ \bar{\omega}_0^2} u^2\right) -\chi \gamma^2 v^2 u^2=0.
\label{disp-fipsi-u}
\end{equation}
We look for a series solution in $u$: 
$$
u=u_0+\delta\, u_1+\delta^2\, u_2+\delta^3\, u^3+\ldots
$$
As a consequence, we expect 
$$
k=\frac{1}{\delta}\, u_0+\left(u_1-\frac{\omega}{v}\right) +\delta\, u_2+\delta^2\, u_3+\ldots
$$
We find at the zeroth order two vanishing degenerate solutions $u_0=0$, to be associated with the modes $H, B$, and also 
\begin{align}
u_{0P} &:= \frac{\bar{\omega}_0}{\gamma v} \sqrt{1-\chi \gamma^2 \frac{v^2}{c^2}},\\
u_{0N} &:= -u_{0P}. 
\end{align} 
Corrections for the two non-degenerate solutions in the first order are 
\begin{align}
u_{1P} &:= -\frac{\omega}{v} \chi \gamma^2 \frac{v^2}{c^2} \frac{1}{1-\chi \gamma^2 \frac{v^2}{c^2}},\\
u_{1N} &:=u_{1P}. 
\end{align} 
Corrections at the first order for the two degenerate zero roots $u_0=0$ arise from the second-order contribution to the 
dispersion relation (as the first order contribution vanishes identically for $u_0=0$): one finds 
\begin{align}
u_{1H} &:=\frac{\omega}{v} \left(1+\sqrt{1+\chi} \frac{v}{c}\right)\frac{1}{1-\chi \gamma^2 \frac{v^2}{c^2}},\\
u_{1B} &:=\frac{\omega}{v} \left(1-\sqrt{1+\chi} \frac{v}{c}\right)\frac{1}{1-\chi \gamma^2 \frac{v^2}{c^2}}. 
\end{align} 
We are not interested in further corrections. It is nice to point out that the expressions we have found are compatible with 
the WKB behaviour of the solutions found in \cite{master}.

\subsection{Scattering coefficients}
We proceed in the same way as exposed in Sections \ref{sec_VarMethod} and \ref{sec_BoundaryConditions}, performing a perturbative expansion of Eq. (\ref{eq_FinalEquation_phipsi}) in the parameter $\eta$. We just give the results of the scattering coefficients, analogous to (\ref{Eq_Coefficients}):
\begin{align}\label{Eq_Coefficients_phipsi}
C_1&= \frac{1}{1+i \eta \tau}\,, \\ 
C_2&= \frac{i \eta}{1+i \eta \tau} \frac{ \pi  (k_P^2-\omega^2 )(-1)^{i(k_H-k_P)/(2\tilde\beta)} }{2 \tilde\beta ^2
	V^2 \epsilon ^2 (k_P-k_N) (k_P-k_B)} \text{csch}\left(\frac{\pi \gamma (k_H-k_P)}{2 \tilde\beta }\right) \,, \\
C_3&= \frac{i \eta}{1+i \eta \tau} \frac{ \pi  (k_N^2-\omega^2 )(-1)^{i(k_H-k_N)/(2\tilde\beta)} }{2 \tilde\beta ^2
	V^2 \epsilon ^2 (k_N-k_P) (k_N-k_B)} \text{csch}\left(\frac{\pi \gamma (k_H-k_N)}{2 \tilde\beta }\right)  \,, \\
C_4&= \frac{i \eta}{1+i \eta \tau} \frac{ \pi  (k_B^2-\omega^2 )(-1)^{i(k_H-k_B)/(2\tilde\beta)} }{2 \tilde\beta ^2
	V^2 \epsilon ^2 (k_B-k_P) (k_B-k_N)} \text{csch}\left(\frac{\pi \gamma (k_H-k_B)}{2 \tilde\beta }\right)  \,,\\
\tau &= \frac{ (k_H^2-\omega^2 ) }{\tilde\beta V^2 \gamma \epsilon ^2 (k_H-k_P) (k_H-k_N)(k_H-k_B)} \,.
\end{align}
Recalling that, from (\ref{disp-fipsior}),
\begin{align}
(k_J^2-\omega^2) = \frac{\chi \omega_0^2\, \gamma^2 \, (\omega+ V k_J)^2}{\omega_0^2-\gamma^2 (\omega+ V k_J)^2} \overset{\omega_{lab}^2 \ll \omega_0^2}{\approx} \chi \gamma^2 (\omega+ V k_J)^2 \,, 
\end{align}
we can see that the expressions of the coefficients (\ref{Eq_Coefficients_phipsi}) basically reduce to (\ref{Eq_Coefficients}) at low frequencies. This fact was expected since the Cauchy dispersion relation in an approximation of the Sellmeier for $\omega_{lab}^2 \ll \omega_0^2$, but it can be viewed also as a check of consistency and robustness of our results.

\section{Comparison with the Orr-Sommerfeld approach}

\begin{figure}
	\centering
	\includegraphics[scale=0.45]{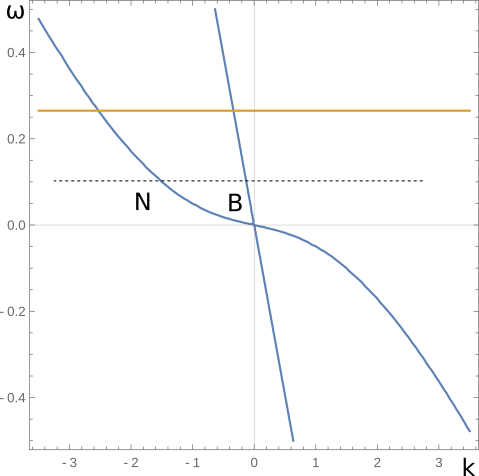}
	\caption{\label{Fig_DRcrit} The dispersion relation (\ref{Eq_nPerturbed}) represented at the peak of the background field, for $\eta<-\eta_{\text{min}}$. In this configuration, for any $\omega$, the modes $H$ and $P$ become imaginary as they experience an event horizon: this is referred to as \emph{transcritical regime}.  }
\end{figure}

Equations like (\ref{Eq_FourthOrder}) are called of generalized Orr-Sommerfeld type. Such equations emerge often in Analogue Gravity and have been studied extensively in \cite{master}. Here the authors developed a general technique for computing the Hawking spectrum in the \emph{transcritical} case, using a perturbative approach in the low-dispersion parameter. 
In this Section we compare our results to those derived with the Orr-Sommerfeld approach: in particular, we establish a relation between the effective temperature we have defined in Eq.~(\ref{Eq_TH}) for the subcritical regime, with the Hawking temperature the authors find in \cite{master} for the transcritical case.  
From equations (\ref{CauchyDisp}) and (\ref{Eq_MuShift})  we we can identify the low-dispersion limit as
\begin{align}
\epsilon^2 := \frac 1 {g^2} \rightarrow 0 \,,\quad \frac {\mu^2} {g^2} =:\theta= \text{const}\,,\quad \frac {\eta}{g^2}=:\rho=\text{const}\,.
\end{align}
Notice that this means that $\mu^2\propto \epsilon^{-2}$ and $\eta\propto \epsilon^{-2}$. 
By eliminating the third order term, Eq. (\ref{Eq_FourthOrder}) can be written as
\begin{align}
&\epsilon^2 f^{(4)}(x) + p_3(x)f''(x) + p_2(x) f'(x) + p_1(x)f(x)=0\,,
\end{align}
where
\begin{eqnarray}
p_3(x) &=& \frac{ \left(g^2 -\mu^2 V^2 +\eta V^2 \text{sech}^2(\tilde\beta x)\right)}{g^2 V^4 \gamma^2}=\frac{ \left(1 -\theta V^2 +\rho V^2 \text{sech}^2(\tilde\beta x)\right)}{ V^4 \gamma^2}\\
p_2(x) &=& \frac{2 i \left(g^2 \left(V^2-1\right) \omega-4\tilde\beta \eta 
	V^3 \tanh (\tilde\beta x) \text{sech}^2(\tilde\beta x)\right)}{g^2 V^5 \gamma^2} \cr
	&=& \frac{2 i \left(\left(V^2-1\right) \omega-4\tilde\beta \rho 
	V^3 \tanh (\tilde\beta x) \text{sech}^2(\tilde\beta x)\right)}{V^5 \gamma^2}\,,\\
p_1(x) &=& \frac{ \left(-2 \tilde\beta^2 \eta V^4 \text{sech}^4(\tilde\beta x)+4 \tilde\beta^2 \eta V^4 \tanh ^2(\tilde\beta x) \text{sech}^2(\tilde\beta x)-g^2 \left(V^2-1\right)^2 \omega^2\right)}{g^2 V^6
	\gamma^2}\cr
	&=& 
	\frac{ \left(-2 \tilde\beta^2 \rho V^4 \text{sech}^4(\tilde\beta x)+4 \tilde\beta^2 \rho V^4 \tanh ^2(\tilde\beta x) \text{sech}^2(\tilde\beta x)- \left(V^2-1\right)^2 \omega^2\right)}{V^6
	\gamma^2}\,.
\end{eqnarray}
Notice that the functions $p_j(x)$ do not depend on $\epsilon$, as in the Corley model. 
This form is directly comparable with the form in \cite{master}. The position of the horizon is defined by
\begin{align}
p_{3}(x) =0 \,.
\end{align}
By changing variable to $z=-e^{2\tilde \beta x}$ as before, we find 
\begin{align}
z_{H\pm} &= \frac{g^2+2\eta V^2-\mu^2 V^2\pm 2\sqrt{g^2 \eta V^2+\eta^2 V^4-\eta \mu^2 V^4}}{g^2-\mu^2 V^2}\cr
&=\frac{1+2\rho V^2-\theta V^2\pm 2\sqrt{\rho V^2+\rho^2 V^4-\rho \theta V^4}}{1-\theta V^2}\,.
\end{align}
The corresponding points $x_{H\pm}$ are real if 
\begin{align}
\Big\{\eta < -\frac {(g^2-\mu^2 V^2)}{V^2} = : -\eta_{\text{min}}\Big\} \cup \Big\{\eta > 0 \Big\}\,.
\end{align}
In this comparison with the critical regime, we just consider the case of positive Kerr nonlinearity: $\delta n>0\implies \eta<0$. Thus we neglect positive values of $\eta$, and we identify $\eta<-\eta_{\text{min}}$ as the condition of transcriticality of the perturbation. Notice that $\eta_{\text{min}}>0$. In Figure \ref{Fig_DRcrit} we plot the dispersion relation in the critical case. 
\\
In \cite{master} thermal Hawking radiation was predicted with a temperature
\begin{align}
T_H = \frac{\kappa}{2\pi} = \frac{\gamma^2 V^2 n'(x_H)}{2\pi}\,,
\end{align}
where $n'(x_H)$ is the derivative of the refraction index at the horizon. We compute $n'$ from the Cauchy dispersion formula 
including the background correction in the 
comoving frame: 
\begin{equation}\label{Eq_nPerturb}
n^2(x)=n_0^2-\frac{\lambda}{2 g^2} \psi^2_B( x),
\end{equation}
where 
\begin{equation}
n_0^2:=\frac{\mu^2} {g^2}\,.
\end{equation}
Here we do not consider the dependence on $\omega_{lab}$ of the refractive index (Eq. (\ref{Eq_nPerturbed})), because the Orr-Sommerfeld approach is true in the low-dispersion limit ($\omega_{lab}\sim 0$): indeed, the event horizon is defined by
\begin{align}\label{Eq_HorizonOrrSomm}
p_3(x_H)=0 \quad \iff\quad  n^2(x_H)=\frac 1 {V^2} \,.
\end{align} 
With these considerations, we find
\begin{align}
n'(x_H) = \frac V {2}\left(n^2(x_H)\right)'= V \left(-\frac{\lambda}{2 g^2}\psi_B^2(x_H)\right)\frac{\psi_B'(x_H)}{\psi_B(x_H)}\,,
\end{align}
and using (\ref{Eq_nPerturb}), (\ref{Eq_HorizonOrrSomm}) and (\ref{Eq_Background}), we and up with
\begin{align}\label{Eq_TH_critic}
T_H =& \frac{\beta\gamma V}{2\pi}\frac{g^2-\mu^2V^2}{g^2}\frac{1+z_H}{1-z_H}\cr
=& \frac{\beta\gamma V}{2\pi}\frac{g^2-\mu^2V^2}{g^2}\sqrt{\frac{|\eta| -\eta_{\text{min}}}{|\eta|}} \,.
\end{align}

This result is surprisingly similar to (\ref{Eq_TH}), although they were computed in very different ways, one within the critical regime and the other in the subcritical.  We notice that if the expression (\ref{Eq_TH}) for $T_{pert}$ is computed for near critical velocity $$ n_0^2 =\frac {\mu^2} {g^2}\lesssim \frac 1 {V^2}\,,$$ and compare it to (\ref{Eq_TH_critic}) in the very critical case ($|\eta|\gg\eta_{\text{min}}$), we find 
\begin{align}
T_{pert}\approx \frac{T_H} 3\,.
\end{align}
We are not able to give an interpretation of the missing factor $3$ with respect to the case of real turning points. We may suggest that it could be an effect of considering a non-monotonic background such as (\ref{Eq_Soliton}), whereas the results of the Orr-Sommerfeld approach in \cite{master} were computed assuming a monotonic background. In any case, the similarity of the critical and subcritical temperatures is interesting and it suggests that they are related by some physical mechanism: this will be a matter for future studies.

The same estimations can be done using the $\phi\psi$ model, introduced in Section \ref{sec_PhiPsi}. We set $c=1$. The temperature estimated from the coefficients (\ref{Eq_Coefficients_phipsi}) is
\begin{align}\label{Eq_TH_PhiPsi}
\beta_{pert} = \frac 1 {T_{pert}} = \frac{2\pi \gamma}{2\beta} \lim\limits_{\omega\rightarrow 0} \frac{2 k_H - k_P - k_N}{\omega} = \frac{2 \pi  \left(1+\gamma ^2 V^2 \chi +V \sqrt{\chi +1}\right)}{\beta  \gamma  V \left(1-V^2 (\chi +1)\right)} \,.
\end{align}
For near-critical velocity
\begin{align}\label{Eq_NearCriticalApprox}
n_0^2 =1+\chi \lesssim \frac 1 {V^2}\,,
\end{align}
we see that the term between brackets in the numerator of (\ref{Eq_TH_PhiPsi}) is 
\begin{align}
\left(1+\gamma ^2 V^2 \chi +V \sqrt{\chi +1}\right) \approx 3\,,
\end{align}
which gives
\begin{align}
T_{pert} \approx \frac {\beta  \gamma  V \left(1-V^2 (\chi +1)\right)}{6 \pi} \,.
\end{align}
The analysis of the critical case with the Orr-Sommerfeld approach gives that criticality is found for
\begin{align*}
\eta < -\eta_{\text{min}}\,, \quad \eta_{\text{min}}:= \frac 1 \chi (1 - \chi\gamma^2 V^2)>0\,
\end{align*}
and  the Hawking temperature is
\begin{align}
T_H 
=& \frac{\beta }{2\pi \chi V \gamma}(1- V^2(1+\chi)) \sqrt{\frac{|\eta| - \eta_{\text{min}}}{|\eta|}}  \cr 
\approx& \frac{\beta \gamma V }{2\pi}(1- V^2(1+\chi)) \sqrt{\frac{|\eta| - \eta_{\text{min}}}{|\eta|}}\,,
\end{align}
where in the last line we used (\ref{Eq_NearCriticalApprox}) as before. We see that, once again, the relation between the two temperatures (for $|\eta|\gg\eta_{\text{min}}$) is
\begin{align}
T_{pert} \approx \frac{T_H}{3}\,,
\end{align}
confirming what we found using the Cauchy model. 

\section{Conclusions}

 We proposed a new approach to Hawking-like radiation in the subcritical case for a particular, but arguably realistic, class of soliton-like backgrounds in a nonlinear dielectric. The method allows a straightforward analytical solution of the scattering problem at the leading perturbation order of the amplitude of the background field, represented by the parameter $\eta$. With respect to other existing techniques, our approach does not rely on the approximation of weak dispersion, and indeed our predictions are not restricted to the $\omega\sim 0$ region.
We tested our approach on a simplified model of scalar electrodynamics and checked its robustness on the scalar field reduction of the Hopfield model, which strongly corroborates the results obtained in the simplified model. In both cases, we can define an effective temperature associated with the spectrum of the emitted radiation, which, in the limit where the subcritical case approaches the transcritical one, is one-third of the Hawking temperature estimated by other established methods for the transcritical regime. The interpretation of this fact is not yet available to us: future work will be focused on the transition to the critical regime, and it will possibly give us a clearer view. 
 
The perturbative expansion we propose, although being naturally suited for the study of the subcritical regime, is not theoretically limited to this case: the problem of the transition to the transcritical regime is configured, in this context, as possibly a matter of being able to compute enough perturbative orders. 
This can be corroborated by the fact that a new mathematical perspective on the phenomenon of particle creation was provided, as we related particle creation to the solution of a fourth-order Fuchsian equation. Fuchsian equations have many well-studied properties and much of that theoretical machinery may be applied to the problem of analogue Hawking radiation, potentially extending our analytical comprehension of the phenomenon. This will be a matter for future studies.

\newpage
\appendix

\section{The conserved current}\label{App_Current}
We compute the current vector, evaluated on normal modes $e^{i k_\mu x^\mu}$ in a generic frame moving with velocity $V$ with respect to the laboratory. From Eq. (\ref{EOM_LinCauchy}) we find that if $\psi(x) = e^{i k_\mu x^\mu}$, then $\phi(x)= -i\frac{g n^\mu k_\mu}{(v^\mu k_\mu)^2} e^{i k_\mu x^\mu}$: substituting these fields into (\ref{cov_current}) we find
\begin{align}
J^\mu \propto v^\mu\, \omega_{lab}\left(1+\frac{\mu^2 + \omega_{lab}^2}{\omega_{lab}^2} \right) + n^\mu \, \frac{g^2 k_{lab}}{\omega_{lab}^2}\,,
\end{align}
where $v^\mu=(\gamma, -V\gamma)$, $n^\mu = (-\gamma V, \gamma)$, $\omega_{lab}=v^\mu k_\mu=\gamma(\omega + V k)$, $k_{lab}=-n^\mu k_\mu =\gamma(k + V\omega)$. 
We can verify that $J^\mu$ is time-like. Indeed,
\begin{align}
J^\mu J_\mu = \omega_{lab}^2\left(1 + \frac{2(\mu^2 + \omega_{lab}^2)}{\omega_{lab}^2} + \frac{(\mu^2 + \omega_{lab}^2)^2}{\omega_{lab}^4}  - \frac{g^2(\mu^2 + \omega_{lab}^2)}{\omega_{lab}^4} \right) \,,
\end{align}
which is positive provided that $  \mu^2 > g^2 $, which is the condition we required at the beginning (see below Eq. (\ref{CauchyDisp})). Thus $J^\mu$ is time-like, so the sign of $J^0$ is constant in every inertial reference frame: since in the lab frame $\text{sign}(J^0) = \text{sign}(\omega_{lab})$, this must be true in every frame.  

We now define the dispersion relation function
\begin{align}\label{Eq_DR}
\text{DR}(\omega, k) :=\mu^2 + \omega_{lab}^2- \frac{g^2 k_{lab}^2}{ \omega_{lab}^2 } \,.
\end{align}
The free normal modes of the theory satisfy $DR(\omega,k)=0$. It is easy to show that the following relations hold
\begin{align}
J^0 &\propto \partial_\omega \text{DR}|_{\text{DR}=0} \,,\\
J^x &\propto -\partial_k \text{DR}|_{\text{DR}=0} \,.
\end{align}
Thus, we can identify the measure of the Hilbert space of free normal modes as 
\begin{align}\label{eq_measureHilbert}
d\mu_j := \frac{d\omega}{2\pi \partial_k \text{DR}|_{k=k_j(\omega)}}\,,
\end{align} 
where we are writing the field theory in the frequency representation and $k_j$ are the different \textit{real} solutions of $\text{DR}=0$. It is also straightforward to show
\begin{align}
	J^x = v_g(\omega) \,J^0 \,,
\end{align}
where $v_g(\omega)=\left.\frac{d\omega}{dk}\right|_{\text{DR}=0}$ is the group velocity of the normal modes. This relation could be derived also from the implicit function theorem, which states $\frac{\partial_k \text{DR}}{\partial_\omega \text{DR} } =- \left.\frac{d\omega}{dk}\right|_{\text{DR}=0}$.

We can write the flux $J^x$ also in a more convenient way for future computations. Since we called $k_H(\omega)$, $k_P(\omega)$, $k_N(\omega)$ and $k_B(\omega)$ the four solutions of the dispersion relation, we can write Eq. (\ref{Eq_DR}) as
\begin{align}
\text{DR}(\omega, k) = \frac{\gamma^2 V^4}{\omega_{lab}^2}\prod_{j=H,P,N,B}(k-k_j(\omega))\,.
\end{align}
It is now easy t verify that
\begin{align}
\partial_\omega  \text{DR}|_{k=k_i(\omega)} = -\frac{\gamma^2 V^4}{\omega_{lab}^2|_{k_i}} \frac{\partial k_i}{\partial \omega} \prod_{j\neq i} (k_i(\omega) - k_j(\omega))\,,
\end{align}
and then
\begin{align}\label{Eq_smartfluxes}
J^x(e^{-i\omega t + i k_i(\omega) x}) = v_i(\omega)\partial_\omega  \text{DR}|_{k_i}= -\frac{\gamma^2 V^4}{\omega_{lab}^2|_{k_i}}  \prod_{j\neq i} (k_i(\omega) - k_j(\omega))\,.
\end{align}

\section{The conservation law}\label{App_ConservationLaw}
To understand the conservation law (\ref{Eq_CurrentConservation}), we must consider that the initial and final states of the scattering should be \textit{wave packets} rather than plane waves. A normalized wave packet centred around the frequency $\omega$ and momentum $k_j(\omega)$ (here $k_j(\omega)$ is any of the solutions of the dispersion relation) has the form
\begin{align}\label{Eq_wavepacket}
\int \frac{d\omega}{2 \pi \partial_k DR |_{k_j(\omega)}} f_\epsilon(\omega)\,e^{i\omega t -i k_j(\omega) x} \,,
\end{align} 
where $\frac{d\omega}{2 \pi \partial_k DR |_{k_j(\omega)}}$ is the measure of the Hilbert space defined by plane waves (in the frequency representation) and $f_\epsilon(\omega)$ is a distribution centred around $\omega$ with a small bandwidth $\epsilon$, which satisfies
\begin{align*}
\int \frac{d\omega}{2 \pi \partial_k DR |_{k_j(\omega)}} |f_\epsilon(\omega)|^2 = 1\,.
\end{align*}
We can see Eq. (\ref{Eq_wavepacket}) represents a wave packet by evaluating it at $x=v t $, for some velocity $v$: for $t=\pm \infty$ we can say that the dominant contribution comes from the stationary point, which is $$1-V\frac{\partial k_j}{\partial \omega}=0 \,,$$ which means that the function (\ref{Eq_wavepacket}) is travelling at a constant velocity $v=v_j(\omega)$, where $v_j(\omega)$ is the group velocity of the mode $k_j$. 

The asymptotic plane-wave solution (\ref{Eq_AsymptoticBehaviour}) corresponds to the wave packets
\begin{align}\label{eq_asymptoticWavePackets}
\psi(x,t) \sim \int \frac{d\omega}{2 \pi \partial_k DR |_{k_H(\omega)}} f_\epsilon(\omega) \,e^{i\omega t} \begin{cases}
e^{-i k_H x} + C_2 e^{-i k_P x} + C_3 e^{-i k_N x} + C_4 e^{-i k_B x}\,, & x\rightarrow -\infty \,,\\
C_1 e^{-i k_H x}\,, & x\rightarrow +\infty \,.
\end{cases}
\end{align}
At $t=-\infty$ the only contribution comes from the in-going $H$-mode, so the measure of integration is chosen such that $$\left(\psi|\psi\right)_{t=-\infty}=1 \equiv \int dx J^0(\psi(x,t=-\infty))\,.$$
The current conservation implies
\begin{align}
\frac d {dt} \int dx J^0 = \int dx \partial_0J^0 = \int dx \partial_x J^x =0\,.
\end{align}
At $t=+\infty$ the solution splits into four outgoing localized wave packets with central momenta: each mode propagates with a different group velocity, so we can assume that the wave packets are not overlapping at $t=+\infty$. With these considerations, the computation of the norm gives
\begin{align}
 &\int \frac{d\omega}{2 \pi \partial_k \text{DR}|_{k_H(\omega)}} \frac{d\omega'}{2 \pi \partial_k \text{DR}|_{k_H(\omega')}} f_\epsilon(\omega) f_\epsilon^*(\omega')  \sum_{i=H,P,N,B}|C_i|^2 2\pi\partial_\omega \text{DR}|_{k_i} \delta(k_H(\omega)-k_H(\omega')) =\cr  
 &= \int \frac{d\omega}{2 \pi \partial_k \text{DR}_{k_H}}|f_\epsilon(\omega)|^2 \frac{1}{2 \pi \partial_k \text{DR}_{k_H}}  \sum_{i=H,P,N,B}|C_i|^2 2\pi v_i \partial_\omega  \text{DR}|_{k_i} \equiv 1 \,.
\end{align}
The last equality comes from current conservation. By recalling $\partial_k DR |_{k_j(\omega)} =- v_j(\omega)\partial_\omega DR |_{k_j(\omega)}$ (from the implicit function theorem) we see that the sign of each term is determined by the product $v_j(\omega)\,\omega_{lab}$. Since the equality must hold for every distribution $f_\epsilon$, we finally obtain the following relation between the absolute values,
\begin{align}
1 = |C_1|^2 + |C_2|^2 \left|\frac{v_P\partial_\omega DR|_{k_P}}{v_H\partial_\omega DR|_{k_H}}\right| - |C_3|^2 \left|\frac{v_N\partial_\omega DR|_{k_N}}{v_H\partial_\omega DR|_{k_H}}\right| + |C_4|^2 \left|\frac{v_B\partial_\omega DR|_{k_B}}{v_H\partial_\omega DR|_{k_H}}\right| \,,
\end{align}
which is precisely (\ref{Eq_CurrentConservation}).

\section{Solution of first order equation}\label{app_Solution}

We start computing the first-order solution $u_1(z)$ (Eq. (\ref{Eq_u1_formal})) around $z=0$ ($x=-\infty$). We proceed as exposed in Section \ref{sec_VarMethod} to derive the expression of the first-order solution. In the scattering of the $H$-mode, we expect the asymptotic coefficient of the $H$-mode to be $O(1)$, while all the others should be $O(\eta)$ (see Section \ref{sec_BoundaryConditions}): for this reason, the zero-order solution we put into the source term (\ref{Eq_SourceTerm}) corresponds to a single $H$-mode, which means
\begin{align}
u_0(z)= z^{i\alpha_H} \,,\quad  i \alpha_H= i\frac {k_H} {2\beta} +\frac 3 2 + i\frac \Omega V \,.
\end{align}
From this choice, we get
\begin{align}
r_1(z) &= \frac{z^{-3+i \alpha_H} \left(4 \alpha_H^2 (z-1)^2+4 i \alpha_H \left(5 z^2-6 z+1\right)-25 z^2+2 z-1\right)}{4 (z-1)^4 V^2 y^2} \,, \\
\frac{W_1(z)}{W(z)}&= \frac{i \left(4 \alpha_H^2 (z-1)^2+4 i \alpha_H \left(5 z^2-6 z+1\right)-25 z^2+2 z-1\right)}{4 (z-1)^4 V^2 \gamma^2 (\alpha_H-\alpha_P)
   (\alpha_H-\alpha_N) (\alpha_H-\alpha_B)} \,,\\
\frac{W_2(z)}{W(z)}&= -\frac{i \left(4 \alpha_H^2 (z-1)^2+4 i \alpha_H \left(5 z^2-6 z+1\right)-25 z^2+2 z-1\right) z^{i (\alpha_H-\alpha_P)}}{4 (z-1)^4 V^2 \gamma^2
   (\alpha_P-\alpha_H) (\alpha_P-\alpha_N) (\alpha_P-\alpha_B)} \,, \\
\frac{W_3(z)}{W(z)}&=-\frac{i \left(4 \alpha_H^2 (z-1)^2+4 i \alpha_H \left(5 z^2-6 z+1\right)-25 z^2+2 z-1\right) z^{i (\alpha_H-\alpha_N)}}{4 (z-1)^4 V^2 \gamma^2
   (\alpha_N-\alpha_H) (\alpha_N-\alpha_P) (\alpha_N-\alpha_B)} \,, \\
\frac{W_4(z)}{W(z)}&= -\frac{i \left(4 \alpha_H^2 (z-1)^2+4 i \alpha_H \left(5 z^2-6 z+1\right)-25 z^2+2 z-1\right) z^{i (\alpha_H-\alpha_B)}}{4 (z-1)^4 V^2 \gamma^2
   (\alpha_B-\alpha_H) (\alpha_B-\alpha_P) (\alpha_B-\alpha_N)} \,.
\end{align}
Computing the integrals in (\ref{Eq_u1_formal}), we get
{\small
\begin{align}\label{Eq_u1_explicit}
\tilde u_1(z)&=\frac{i\, z^{i\alpha_H}}{4 V^2
\gamma^2} \left(\frac{-4 \alpha_H^2 (z-1)^2-4 i \alpha_H \left(5 z^2-8 z+3\right)+25 z^2-26 z+9}{(z-1)^3 (\alpha_H-\alpha_P)
		(\alpha_H-\alpha_N) (\alpha_H-\alpha_B)}+\right. \cr 
&\left.\frac{z \left( i (2 \alpha_H+5 i )^2 \, _2F_1(2,i (\alpha_H-\alpha_P)+1,i
		(\alpha_H-\alpha_P)+2;z)+16 (\alpha_H+3 i) \, _2F_1(3,i (\alpha_H-\alpha_P)+1,i (\alpha_H-\alpha_P)+2;z)\right)}{(\alpha_H-\alpha_P) (\alpha_H-\alpha_P-i) (\alpha_P-\alpha_N)
		(\alpha_P-\alpha_B)} \right.\cr
	&\left. -\frac{z\left( 24\, i \, _2F_1(4,i
		(\alpha_H-\alpha_P)+1,i (\alpha_H-\alpha_P)+2;z)\right)}{(\alpha_H-\alpha_P) (\alpha_H-\alpha_P-i) (\alpha_P-\alpha_N)
		(\alpha_P-\alpha_B)}+\right. \cr 
	&\left.\frac{z \left( i (2 \alpha_H+5 i )^2 \, _2F_1(2,i (\alpha_H-\alpha_N)+1,i
		(\alpha_H-\alpha_N)+2;z)+16 (\alpha_H+3 i) \, _2F_1(3,i (\alpha_H-\alpha_N)+1,i (\alpha_H-\alpha_N)+2;z)\right)}{(\alpha_H-\alpha_N) (\alpha_H-\alpha_N-i) (\alpha_N-\alpha_P)
		(\alpha_N-\alpha_B)} \right.\cr
	&\left. -\frac{z\left( 24\, i \, _2F_1(4,i
		(\alpha_H-\alpha_N)+1,i (\alpha_H-\alpha_N)+2;z)\right)}{(\alpha_H-\alpha_N) (\alpha_H-\alpha_N-i) (\alpha_N-\alpha_P)
		(\alpha_N-\alpha_B)}+\right.\cr 
	&\left.\frac{z \left( i (2  \alpha_H+5 i)^2 \, _2F_1(2,i (\alpha_H-\alpha_B)+1,i
		(\alpha_H-\alpha_B)+2;z)+16 (\alpha_H+3 i) \, _2F_1(3,i (\alpha_H-\alpha_B)+1,i (\alpha_H-\alpha_B)+2;z)\right)}{(\alpha_H-\alpha_B) (\alpha_H-\alpha_B-i) (\alpha_B-\alpha_P)
		(\alpha_B-\alpha_N)} \right.\cr
	&\left. -\frac{z\left( 24\, i \, _2F_1(4,i
		(\alpha_H-\alpha_B)+1,i (\alpha_H-\alpha_B)+2;z)\right)}{(\alpha_H-\alpha_B) (\alpha_H-\alpha_B-i) (\alpha_B-\alpha_P)
		(\alpha_B-\alpha_N)}\right) \,.
\end{align}}
This is a \emph{particular} solution of (\ref{SetEquations}): the general expression is obtained by adding a linear combination of free-field solutions (with coefficients of order $O(\eta)$). The coefficients of such combinations will be determined later based on the boundary conditions. From (\ref{Eq_u1_explicit}) we can compute the asymptotic expression of $\tilde u_1$ around $z=0$ ($x=-\infty$):
\begin{align}
\tilde u_1(z\approx 0)&= c_H \,z^{i \alpha_H}\left(1 + O(z)\right)\,, \cr
c_H &= -\frac{i \left(-4 \alpha_H^2-12 i \alpha_H+9\right)}{4 V^2 \gamma^2 (\alpha_H-\alpha_P) (\alpha_H-\alpha_N) (\alpha_H-\alpha_B)} \,.
\end{align}
The function (\ref{Eq_u1_explicit}) is defined on the whole complex plane: to write the asymptotic expression at $z=\infty$ ($x=+\infty$) we use the connection formulas of the hypergeometric function:
\begin{align}
 _2F_1(a,b,c;z) =&\frac{\left(-z\right)^{-a} \Gamma (c) \Gamma (b-a) }{\Gamma (b) \Gamma
	(c-a)} \, _2F_1\left(a,a-c+1;a-b+1;\frac 1 z\right) \cr 
&+\frac{\left(-z\right)^{-b} \Gamma (c) \Gamma (a-b)}{\Gamma (a) \Gamma (c-b)}  \, _2F_1\left(b,b-c+1;-a+b+1;\frac 1 z\right) \,.
\end{align}
Expanding around $z=\infty$ we find
\begin{align}
\tilde u_1(z\approx \infty) =& -c_P\, z^{i \alpha_P} \left(1 + O(z)\right) - c_N\, z^{i \alpha_N} \left(1 + O(z)\right) -c_B\, z^{i \alpha_B} \left(1 + O(z)\right) \,,\cr  
c_P &= -\frac{i \pi  (2 \alpha_P+3 i)^2 \left(-1\right)^{i (\alpha_H-\alpha_P)} \text{csch}(\pi 
	(\alpha_H-\alpha_P))}{4 V^2 \gamma^2 (\alpha_P-\alpha_N) (\alpha_P-\alpha_B)}\,, \cr
c_N &=-\frac{i \pi  (2 \alpha_N+3 i)^2 \left(-1\right)^{i (\alpha_H-\alpha_N)} \text{csch}(\pi 
	(\alpha_H-\alpha_N))}{4 V^2 \gamma^2 (\alpha_N-\alpha_P) (\alpha_N-\alpha_B)}\,, \cr
c_B &= -\frac{i \pi  (2 \alpha_B+3 i)^2 \left(-1\right)^{i (\alpha_H-\alpha_B)} \text{csch}(\pi 
	(\alpha_H-\alpha_B))}{4 V^2 \gamma^2 (\alpha_B-\alpha_N) (\alpha_B-\alpha_P)}\,.
\end{align}
From the boundary conditions (\ref{Eq_AsymptoticBehaviour}) we see that we do not have the modes $P$, $N$ and $B$ at right infinity; so we add to the particular solution the linear combination $$ c_P z^{i \alpha_P} +  c_N z^{i \alpha_N} + c_B z^{i \alpha_B} \,.$$
The first-order correction to the solution is thus
\begin{align*}
 u_1(z)=\tilde u_1(z) +  c_P z^{i \alpha_P} +  c_N z^{i \alpha_N} + c_B z^{i \alpha_B} \,,
\end{align*}
 such that $ u_1(z\approx \infty)=0$. Finally, we can write down the solution $u(z)$ at first perturbative order and its asymptotic behaviour:
\begin{align}\label{Eq_u_final}
u(z) \sim \begin{cases}  
(1 + \eta\, c_H )\,z^{i \alpha_H} + \eta\, c_P \,z^{i \alpha_P} + \eta\, c_N \,z^{i \alpha_N} + \eta\, c_B \,z^{i \alpha_B} \,,& z\rightarrow 0 \,,\\
z^{i \alpha_H} \,,& z\rightarrow \infty\,.
\end{cases}
\end{align}
The coefficients (\ref{Eq_Coefficients}) are obtained dividing (\ref{Eq_u_final}) by $(1+\eta\,c_H)$ and recalling $i \alpha_j= i\frac {k_j} {2\beta} +\frac 3 2 + i\frac \Omega V$.



\begin{thebibliography}{99}

\bibitem{weinfurtner-prl} 
  S.~Weinfurtner, E.~W.~Tedford, M.~C.~J.~Penrice, W.~G.~Unruh and G.~A.~Lawrence,
  ``Measurement of stimulated Hawking emission in an analogue system,''
  Phys.\ Rev.\ Lett.\  {\bf 106}, 021302 (2011).

\bibitem{weinfurtner-book}
  S.~Weinfurtner, E.~W.~Tedford, M.~C.~J.~Penrice, W.~G.~Unruh and G.~A.~Lawrence,
  ``Classical aspects of Hawking radiation verified in analogue gravity experiment,''
  Lect.\ Notes Phys.\  {\bf 870}, 167 (2013).

\bibitem{Parentani_numerical}
F.~Michel and R.~Parentani,
``Probing the thermal character of analogue Hawking radiation for shallow water waves?,''
Phys. Rev. D \textbf{90}, no.4, 044033 (2014).

\bibitem{euve} 
  L.~P.~Euv\'e, F.~Michel, R.~Parentani and G.~Rousseaux,
  ``Wave blocking and partial transmission in subcritical flows over an obstacle,''
  Phys.\ Rev.\ D {\bf 91}, no. 2, 024020 (2015).

\bibitem{parentani-sub} 
  S.~Robertson, F.~Michel and R.~Parentani,
  ``Scattering of gravity waves in subcritical flows over an obstacle,''
  Phys.\ Rev.\ D {\bf 93}, no. 12, 124060 (2016).

\bibitem{coutant-subcritical} 
  A.~Coutant and S.~Weinfurtner,
  ``The imprint of the analogue Hawking effect in subcritical flows,''
  Phys.\ Rev.\ D {\bf 94}, no. 6, 064026 (2016).

\bibitem{bremmer49}
H.~Bremmer, 
``The W.K.B. approximation as the first term of a geometric-optical series,''
Comm. Pure Appl. Math. {\bf 4} 105 (1951).

\bibitem{bremmer}
H.~Bremmer, 
``The propagation of electromagnetic waves through a stratified medium and its W.K.B. approximation for oblique incidence,''
Physica {\bf 15}, 593 (1949).


\bibitem{coutant-kdv} 
  A.~Coutant and S.~Weinfurtner,
  ``Low-frequency analogue Hawking radiation: The Korteweg-de Vries model,''
  Phys.\ Rev.\ D {\bf 97}, no. 2, 025005 (2018).

\bibitem{coutant-bdg} 
  A.~Coutant and S.~Weinfurtner,
  ``Low frequency analogue Hawking radiation: The Bogoliubov-de Gennes model,''
  Phys.\ Rev.\ D {\bf 97}, no. 2, 025006 (2018).


\bibitem{fincaru}
S.~Finazzi and I.~Carusotto,
``Quantum vacuum emission in a nonlinear optical medium illuminated by a strong laser pulse,''
Phys. Rev. A \textbf{87}, no.2, 023803 (2013).


\bibitem{finazzi-carusotto} 
  S.~Finazzi and I.~Carusotto,
  ``Spontaneous quantum emission from analog white holes in a nonlinear optical medium,''
  Phys.\ Rev.\ A {\bf 89}, no. 5, 053807 (2014).




\bibitem{master} 
  F.~Belgiorno, S.~L.~Cacciatori and A.~Vigan\`o,
  ``Analog Hawking effect: A master equation,''
  Phys.\ Rev.\ D {\bf 102}, no. 10, 105003 (2020).

\bibitem{bec} 
  F.~Belgiorno, S.~L.~Cacciatori, A.~Farahat and A.~Vigan\`o,
  ``Analog Hawking effect: BEC and surface waves,''
  Phys.\ Rev.\ D {\bf 102}, no. 10, 105004 (2020).


\bibitem{ni}

T. Nishimoto, 
``On the Orr-Sommerfeld type equations, II Connection formulas,''
K\u{o}dai Math. Sem. Rep. 29, (1978), 233.

\bibitem{ni-I}
T. Nishimoto, 
``On the Orr-Sommerfeld type equations, I W.K.B. approximation,''
K\u{o}dai Math. Sem. Rep. 24, (1972), 281.






\bibitem{prd2015}
F.~Belgiorno, S.~L.~Cacciatori and F.~Dalla Piazza,
``Hawking effect in dielectric media and the Hopfield model,''
Phys. Rev. D \textbf{91}, no.12, 124063 (2015).

\bibitem{solitonic}
F.~Belgiorno and S.~L.~Cacciatori,
``Analogous Hawking Effect in Dielectric Media and Solitonic Solutions,''
Universe \textbf{6}, no.8, 127 (2020).

\bibitem{hopfield}
J.~J.~Hopfield, 
``Theory of the Contribution of Excitons to the Complex Dielectric Constant of Crystals,'' 
Phys.~Rev. \textbf{112} (1958), 1555--1567.


\bibitem{Chushev}
A.V. Chueshev, V.V. Chueshev, 
``Variational Formulas of the Monodromy Group
for a Third-Order Equation on a Compact Riemann Surface,''
Journal of Siberian Federal University. Mathematics \& Physics 2022, \textbf{15}(3), 308--318.
	


        \bibitem{hopfield-kerr} 
F.~Belgiorno, S.~L.~Cacciatori, F.~Dalla Piazza and M.~Doronzo,
``Hopfield-Kerr model and analogue black hole radiation in dielectrics,''
Phys. Rev. D \textbf{96}, no.9, 096024 (2017).


\bibitem{hawbook}
F. Belgiorno, S. L. Cacciatori and D. Faccio, \textit{Hawking Radiation}, World Scientific Publishing Company, Singapore (2018).
\end{thebibliography}
\end{document}